\documentstyle[psfig,subfigure]{mn}

\begin{document}
\input{epsf.tex}
\title[Dust Formation In Early Galaxies]
{Dust Formation In Early Galaxies}
\author[H.L.Morgan, M.G.Edmunds]
{H.L.Morgan, $^1$
M.G.Edmunds$^1$ \\
$^1$Department of Physics and Astronomy, University of Wales
\\Cardiff, PO Box 913, Cardiff, CF24 3YB\\}

\maketitle

\begin{abstract}
We investigate the sources and amount of dust in early galaxies.
We discuss dust nucleation in stellar
atmospheres using published extended
atmosphere models, stellar evolution tracks and nucleation
conditions.  The Thermally Pulsating Asymptotic Giant
Branch (TPAGB) phase of intermediate mass stars is likely to be the most
promising site for dust formation in stellar winds.
We present an
elementary model including dust formation timescales in which the
amount of dust in the Interstellar Medium is governed by
chemical evolution. 
The implications of the model for high redshift
galaxies are investigated and we show there is no difficulty in producing
dusty galaxies at redshifts above 5 if supernovae are a dominant source of
interstellar dust.  If dust does not condense efficiently in SNe then
significant dust masses can only be generated at $z > 5$ by galaxies
with a high star formation efficiency.
This is consistent with the high SFR's implied by sub-millimetre
sources found in deep SCUBA surveys. 
We find the visual optical depth for individual star forming clouds
can reach values greater than 1 at very low metallicity (1/100 solar)
provided that the mass-radius exponent of molecular clouds is less
than two.  Most of the radiation from star formation will emerge at IR
wavelengths in the early universe provided that dust is present.  The
(patchy) visual optical depth through a typical early galaxy
will however, remain less than 1 on average until a metallicity of
1/10 solar is reached.
\end{abstract}


\section{Introduction}
\label{intro}
The role of dust in galaxies at present and at earlier epochs is still 
highly controversial.  Recent dust models have attempted to understand
the evolutionary cycle of dust
(Edmunds 2001 (hereafter E01); Dwek 1998 (hereafter Dwek98); Tielens
1998) but there can be no real
advancement whilst there still exists inherent uncertainties over dust
sources, composition, destruction and the dependence of dust
production on metallicity
evolution.  It is well known that dust production in the Interstellar
Medium (ISM) is too slow (timescales for dust growth are smaller than
destructive timescales (Tielens 1998)) so formation sites during stellar 
evolution are required to explain the vast amounts of dust we observe
in the universe.

It is generally accepted that dust is formed in the atmospheres of
evolved stars and ejected via stellar winds with mass loss rates of up 
to $10^{-6}M_\odot yr^{-1}$ (Gehrz 1989; Whittet 1992) and contribute
up to $0.3M_\odot$ of dust to the ISM in our Galaxy per year (Bode
1988).  Others 
however, propose that supernovae explosions must be an important
contributor to dust formation (Dwek \& Scalo 1980, Todini \& Ferrara 2001
(hereafter TF01)) to account for
the dust observed at high redshifts (as seen in Damped Lyman
Alpha systems and in SCUBA data (Frayer \& Scoville 1999)).  The lack of
significant XMM detection of SCUBA sources in a deep field
(Waskett et al, 2003) imply these sources are probably not AGN and
even if they were, then they are likely to be heavily obscured and at
high redshifts ($z>2$).  Almaini et al (1999) also show no significant 
Chandra detections of other sub-mm fields.  If not AGN then they
are probably extremely dusty starburst galaxies, implying that dust is
already present in the high redshift universe.  
It is difficult for the dust to have originated from low-mass evolved
stars at redshifts $\ge 5$ as their 
evolutionary timescales ($10^8$ - $>10^9$yrs) are
comparable to the age of the Universe at that time.  So it is argued then that if dust {\it is} present in the
early universe, then Type II SNe are responsible.  However, some
estimates of the contribution of SNe dust mass to the current ISM show an
injection of only $10^{-4}M_\odot$ per year (Bode 1988; Gehrz 1989), 
and a recent study of the InfraRed emission from three supernova
remnants (SNR) has shown no evidence of dust formation in SN ejecta -
in fact, it was found the emission was probably circumstellar in
origin (Douvion et al 2001a).  
\\
\\
Even dust production in stars has its
own complications. Depending on the star's composition (whether it is carbon
rich (C stars) or oxygen rich (M stars) and what fraction of a stellar 
population is in these stars, there will be a difference in the
returned fraction of carbon and silicate type grains to the ISM.  The 
problem compounds when O rich stars may become C rich as they 
ascend the Asymptotic Giant Branch (AGB) whilst undergoing
pulsations (Marigo et al 1996). IRAS observations (Molster et al 2001) and
observations of AGB stars in the Large Magellanic Cloud (Trams et al
1999) show a high degree of
silicate dust in C stars and it is yet unclear whether this is
due to silicate formation in a previously O rich star or due to an O rich
binary companion (Molster et al 2001).
\\
\\
The questions posed and discussed here are then: when are the
conditions for dust formation reached in stars?  What is the source of 
dust at high redshift and is dust production dependent on metallicity?
How much dust is present in these early galaxies and how will this
affect observations?  The aim of this paper is to ask whether or not
we can build up a significant amount of dust in high redshift galaxies
using the available literature and models.

In Section~\ref{sec:star} we use published stellar atmosphere and
evolution models to determine when dust may form in stars.  We then estimate
the efficiency of dust formation 
in stellar winds and supernovae.  Section~\ref{sec:model} presents the well
established chemical evolution equations and their applications for
following dust evolution in a galaxy.  In Section~\ref{sec:early} we
attempt to understand the source of dust in
early galaxies, using the modified
elementary model from E01.  We incorporate the finite lifetime of
stars into the dust model and compare this with evolutionary
timescales to determine the maximum redshift after which dust can be
observed in the
Universe for different star formation rates and cosmological parameters.
Finally, Section~\ref{sec:conc} gives the conclusions obtained
in this report and an outline of future work.

\section{Dust Production in Stars and SNe}
\label{sec:star}
The majority of work has assumed that the formation of solids from a
gas in stars can be described using classical nucleation
theory.  Condensation of a species X occurs when its
partial pressure in the gas exceeds its vapour pressure in condensed
phase (Whittet 1992), with subsequent particle growth by random
encounters leading to cluster formation. Chemical reactions will
further assist cluster production (Salpeter 1974).  The rate of grain
growth will therefore depend on the temperature, $T_g$ and pressure,
$P_g$ of the gas and the condensation temperature of a species, $T_c$.
The optimum conditions for nucleation are thought to occur in the
range $10^{-8} \leq P_g (Pa) \leq 10^5$ and $T_g \leq 1800K$ (Salpeter 
1974; Yorke 1988).  This
gives very strict restrictions 
on where nucleation can occur and this is thought to be satisfied 
in giant atmospheres at certain stages of evolution.

The type of dust formed will depend on the relative abundance of carbon or
oxygen with carbon dust able to form in temperatures 
$T_g \leq1800K$.  For silicates to form via condensation we require
temperatures below $1500K$ (Bode 1987; Whittet 1992; Salpeter 1974).
These conditions represent the boundary 'nucleation' in
Figures~\ref{atmosa} and (b).

\subsection{Nucleation in Stellar Atmospheres}
\label{nuc}
Unfortunately there is not yet nearly enough hard observational
evidence to follow through the whole process of dust formation in
stars so, in an attempt to investigate when nucleation conditions are
reached in stars, we examine extended atmosphere models for K and M
giants (Plez et al 1992a \& 1992b; Hauschildt et al 1999b;
Brown et al 1993).
We use these models to
estimate the temperature and pressure in the gas out to a stellar
radius where the continuum optical depth at 500nm has fallen to
$10^{-6}$.  The atmospheric conditions depend on the mass of the
star, $M$, its surface gravity ($g={GM \over R^2}$), and its effective
temperature at the surface, $T_{\mathit{eff}}$.  The published models
assume spherical geometry with homogeneous stationary layers in local
thermodynamic equilibrium (LTE) and use numerical codes to
simulate the gas properties in stellar atmospheres
(e.g. {\small{PHOENIX}} (Hauschildt et al 1999a) with these
models available from the website: http://phoenix.physast.uga.edu/NG-giant). 
Although this is probably not a realistic description of conditions in
stellar atmospheres where pulsations, mass loss and inhomogeneities
will surely exist, they will serve to make an estimate of
the mean condition for the processes occuring.

\begin{figure*}
 \subfigure[\label{atmosa}]{\psfig{file=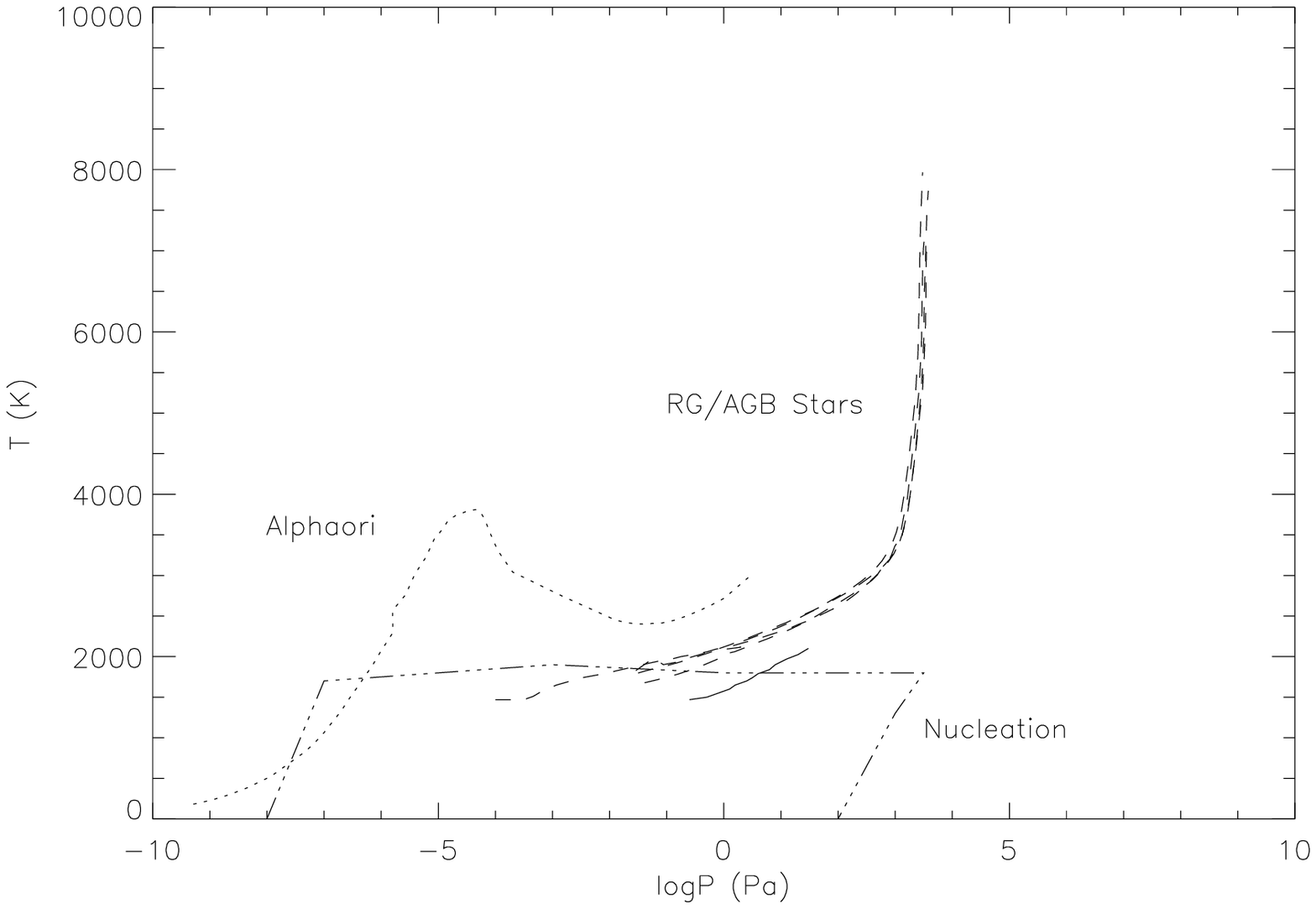,width=8cm,height=6cm}}
 \subfigure[\label{atmosb}]{\psfig{file=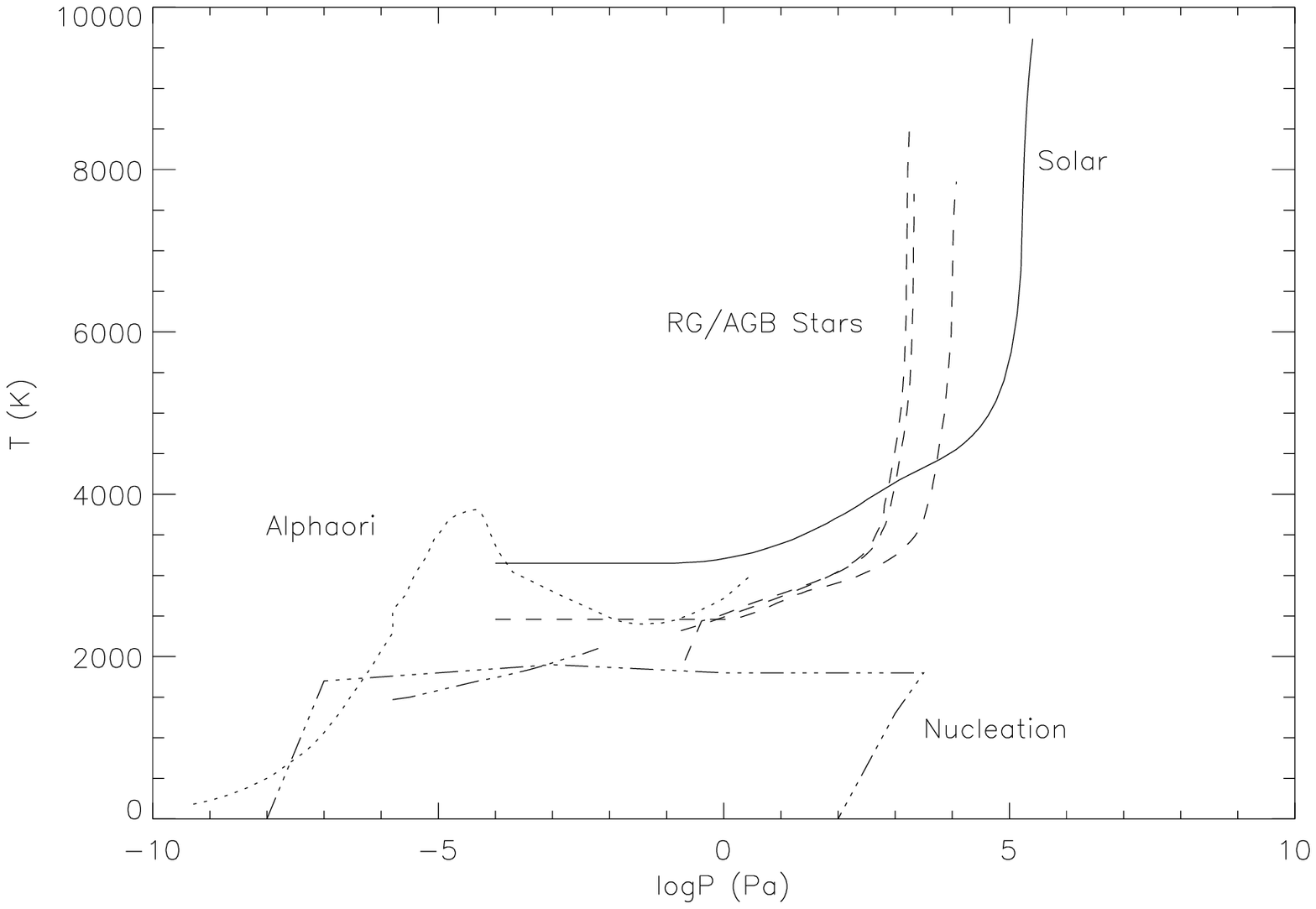,width=8cm,height=6cm}}
 \subfigure[\label{atmosc}]{\psfig{file=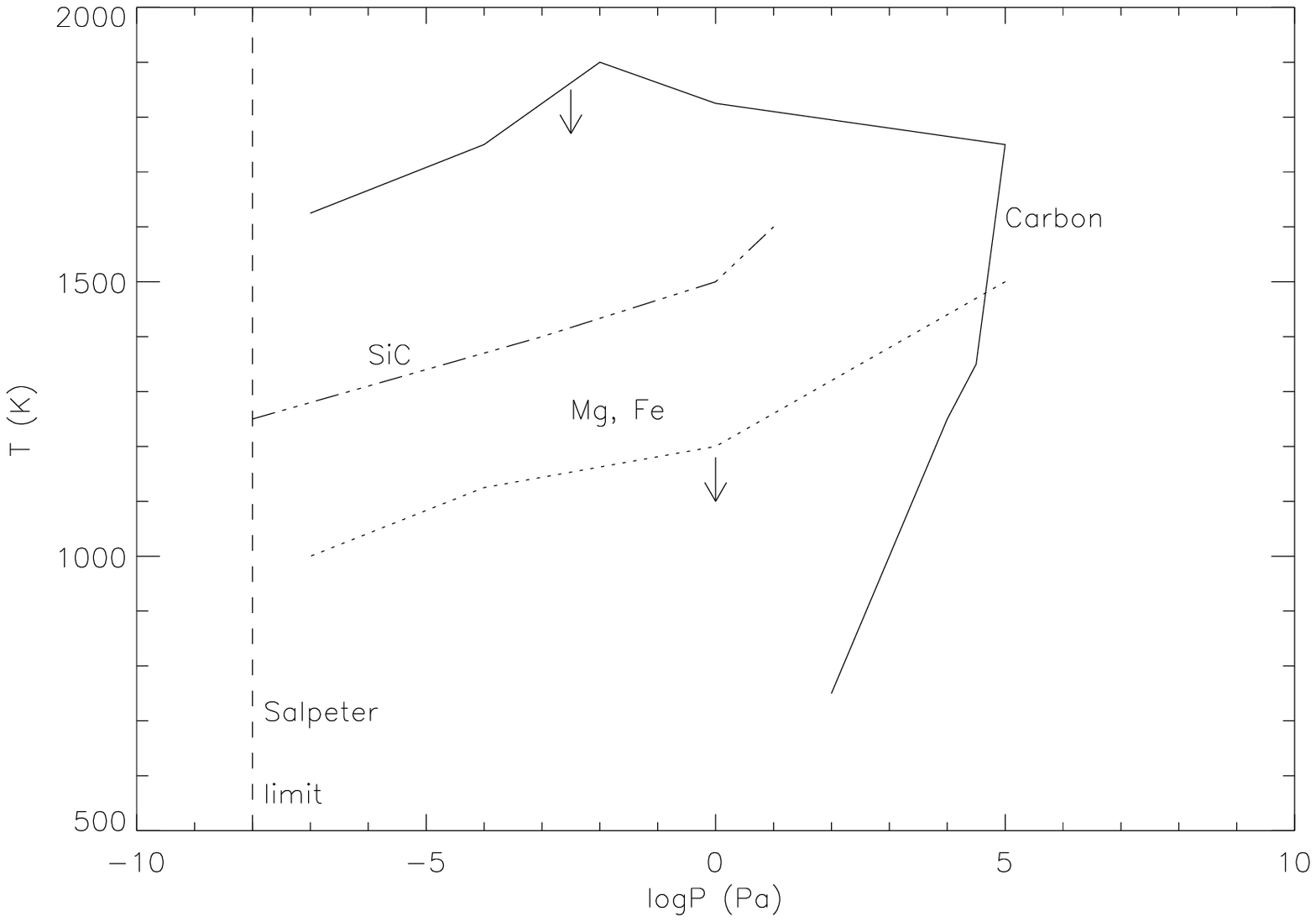,width=8cm,height=6cm}}
 \caption{\small{The gas properties of model atmospheres of different
stars from available literature models (see text).  The RG/AGB curves
represent the temperatures and pressures seen in the atmospheres of
evolved giants if (a) $T_{\mathit{eff}} \leq 3000K$ and (b)
$T_{\mathit{eff}}=3600-3800K$.  The solid curve labelled solar
represents the atmospheric properties for a $1M_{\odot}$ star with
$logT_{\mathit{eff}}=5.5K$ (Hauschildt et al 1999b).  
An empirical atmosphere (Harper et al 2001) representing the giant star
Alpha Ori and the nucleation 
boundary of carbon are included in both figures.  For a higher
effective temperature, the ability to reach the nucleation conditions
in stellar atmospheres decreases (b).  The nucleation conditions in a
carbon-rich gas as
adapted from Yorke (1988) and Salpeter (1974) are shown in (c).  For
an O-rich environment, silicates can be formed if $T_{g}<1500K$ for
the same pressure range.  The
arrows represent the direction in which
nucleation conditions are satisfied for the elements shown.}}
\end{figure*} 

Model atmospheres for C stars were obtained from J\"{o}rgensen et al
(1992) and an
observational model (Harper et al 2001) of the red giant AlphaOri 
is included for comparison (with $T_{\mathit{eff}}=3600K$, $M=12M_\odot$,
$log(g)=0.0$).  The model $P$-$T$ curves are shown in
Figure~\ref{atmosb} for $T_{\mathit{eff}}>3000K$, and
Figure~\ref{atmosa} for $T_{\mathit{eff}}<3000K$. They
clearly show the models do not extend to the low gas
temperatures observed in AlphaOri, and we will assume the boundaries do in
fact extend downwards in a similar manner.  It is also clear that as
$T_{\mathit{eff}}$ increases, the likelihood of the gas temperatures
reaching the nucleation boundary decreases.  The figures in
Hauschildt et al (1999b) show for
various values of $log(g)$, that only those atmospheres with
$T_{\mathit{eff}}<4000k$ will reach gas temperatures below 2000K and hence
attain C nucleation conditions.  In order to reach the silicate nucleation
boundary, we require $T_{\mathit{eff}}\leq 3000K$ (Fig~\ref{atmosc}).

\begin{figure*}
\begin{minipage}{11cm}
\psfig{file=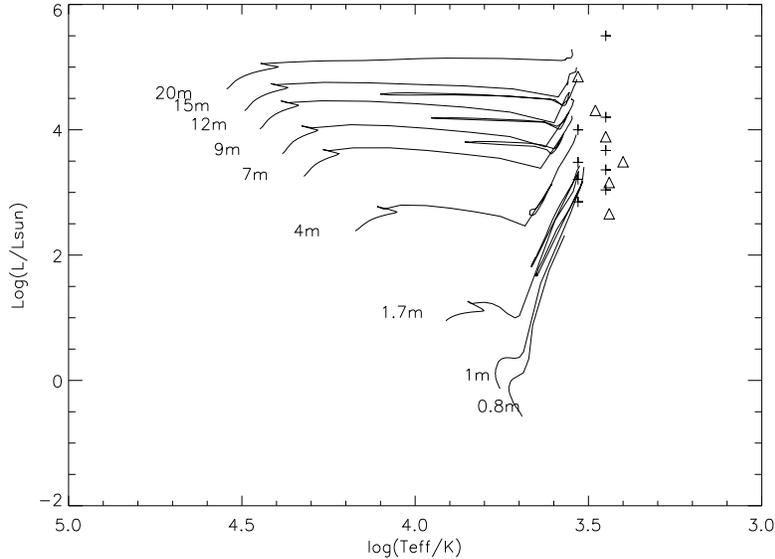,width=11cm,height=8cm}
\end{minipage}
\caption{\small{The HR diagram with added stars corresponding to the model 
atmospheres used.  The triangles represent the K
\& M giants whilst the crosses represent the C stars.}}
\label{hrstars}
\end{figure*}

To determine to which stars these model atmospheres apply, the
luminosity was calculated using Equation~\ref{lum} and simply placed on a
Herstzprung-Russell diagram using their respective $T_{\mathit{eff}}$
values (Fig~\ref{hrstars}).

\begin{equation}
L_*=4{\pi}{\sigma}{T^4_{\em eff}}
\label{lum}
\end{equation}

The stars do indeed correspond to the evolved stages expected, but
for dust nucleation properties to be reached we require the stars to
lie on or just beyond a late AGB phase (evolutionary tracks 
of solar metallicity, $Z_{\odot}=0.02$, from Schaller et al (1992)).  It
appears then that
the stars seen to produce dust using the arguments given here, only
satisfy the
conditions at the very latest stages of their evolution which is only
a very small fraction of their life ($10^4-10^6$years).  This phase
is the TPAGB (Thermally Pulsating Asymptotic Giant Branch) as described
further in Section~\ref{tpagb}.  The stellar evolution models from the
the Padova group (http://pleiadi.pd.astro.it) also give the same conclusion.  The stellar
tracks can reach lower effective temperatures (Pols et al 1995)
depending on the opacitites, overshooting and mixing parameters
used although a recent study of these models suggest there is still
more to be done in this area (Young et al 2001).  We await further
progress on new tracks given by this group using their new stellar
evolution code, {\small{TYCHO}}. 

In further sections we assume that only those stars with 
$T_{\mathit{eff}}<4000K$ and which lie on this evolutionary branch are
significant producers of dust (although this is actually a maximum
limit since these are the conditions for carbon nucleation and we 
would expect to require much lower temperatures for the formation of silicate dust).

It is important to note that because of this reasoning we have ignored dust
production in the atmospheres of stars with $M_i>25M_{\odot}$.
Observationally however, dust shells are seen around a class of
massive stars which suffer large mass loss ($\sim
10^{-5}M_{\odot}yr^{-1}$) in the late stages of their evolution.
These are the Wolf-Rayet stars, which have three phases
characterised by the dominant species of O, C and N in their spectra. 
Dust shells are only seen around the very cool, carbon rich
WR (Dwek98), and will therefore contribute to the production
of carbon dust in the ISM.  Could this be a significant dust formation
site?  Recent observations of a WR binary
system have shown that the formation of dust in these stars may be
increased via wind-wind interactions with at least 20 per cent escaping
and thus enriching the ISM (Marchenko et al 2002).  Their results find
a total dust mass of $2.8\times 10^{-5}M_{\odot}$ which coupled with the mass
loss rate gives
an estimate of the condensation efficiency of dust in WR of $\sim
6$ per cent. But WR stars are also very rare in galaxies as compared to low-intermediate 
mass stars, with the fraction of stars in a given population with
$M>25M_{\odot}$ $\sim 10^{-4}$ implying that the yield of dust from such a
process will not be significant.  

\subsection{Mass Loss and the TPAGB Phase}
\label{tpagb}
The dominant processes in stellar mass loss are 
pulsational and line driven winds with mass loss rates
\footnote[2]{Dust driven winds only become important at high
metallicities where there is sufficient dust production to drive the
gas outwards (Willson 2000).} of $\sim
10^{-4}- 10^{-6} M_{\odot}yr^{-1}$ and
$10^{-7}-10^{-10}{M_\odot}yr^{-1}$.  The important
question to ask is how much of these outflows are in the form of
{\it condensed} material?  Whether or not dust is formed from the gas
will depend on the temperature and the density inside the wind
(Section~\ref{nuc}).  A rough estimate of the dust production in
these winds can be made by investigating condensation conditions in
stellar atmosphere models which in turn are tied to stellar
evolutionary models which predict
$T_{\mathit{eff}}$ and $\dot M$ throughout the stars evolution.

The mass lost throughout the star's lifetime
must satisfy the initial-final mass relationship shown in
Weidermann (1987) and Marigo (2000).
Thus for low-intermediate mass stars $1 \leq M_i\leq 7M_{\odot}$, the
mass lost in stellar winds and novae is the difference between the
initial mass, $M_i$ and the final remnant mass, $M_R$.  For high mass stars
($7<M_i \leq 40M_{\odot}$) the mass is lost via both winds and SNe.
Stars with mass beyond $40M_{\odot}$ may become a black hole and
swallow up the elements produced (Maeder 1992) so they are ignored
here.

\begin{figure*}
\begin{minipage}{11cm}
\psfig{file=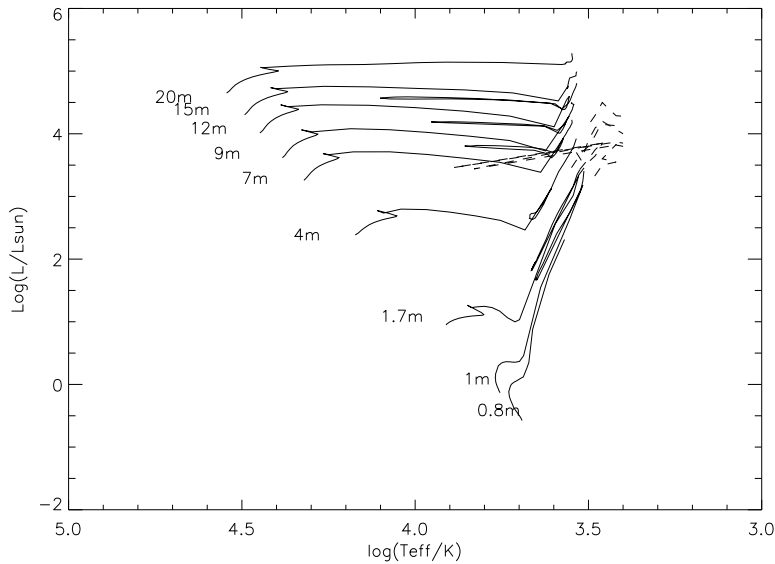,width=11cm,height=8cm}
\end{minipage}
\caption{\small{The evolution of low-intermediate mass stars up to the
end of C burning (solid line) and up through TPAGB phase (dashed line) for
$Z=Z_\odot$.  The PAGB tracks are given in Bl\"{o}cker (1995b), and are not
included here.}}
\label{tpphase}
\end{figure*}

We have followed the evolutionary model grids in Schaller et al
(1992)\footnote[3]{as distributed by the
astronomical data center at NASA Goddard Space Flight Center}, 
up to the end of central C burning for high mass stars,
the early AGB phase for
intermediate stars and up to the He flash for low mass stars.  Some of 
the mass lost during a stars lifetime must occur during a very late
AGB phase or after (the mass lost from the published models does not
explain the difference between the initial mass and remnant
mass\footnote[4]{the points of
greatest mass loss up until the end of the evolutionary grids was
multiplied by the
duration, $\tau$, to give the total mass lost}.  In the TPAGB \& stellar
wind phases, $\dot M$ \& $\tau$ were taken
from Bl\"{o}cker (1995a) and Marigo (1996), but how and where does this
occur? As a star
ascends the AGB its thought that (depending on the previous mass loss
history), it may undergo some He flashes and will pulsate
periodically.  This is the TPAGB phase and will create a dredge up of
material within the star to its surface.  There will be not only a
change in the composition in the atmosphere but also the composition
of the dust (Garcia-Lario \& Pera Calderon 2001; Herwig et al 1999).
The TPAGB phase has mass loss
rates comparable to the AGB phase and lasts for approximately
$10^4-10^6$years depending on the initial stellar mass (7-3$M_\odot$
respectively) and the number of thermal pulses in the
phase (Bl\"{o}cker 1995a).  So there is an opportunity here to create and
inject dust into the ISM, but are the conditions right?  The
conditions for dust formation do appear to be satisfied in the TPAGB
phase as shown in Figure~\ref{tpphase} where the evolutionary tracks of
stars undergoing thermal pulses from Marigo (1996) are added to those
in Schaller et al (1992) for low-intermediate mass stars.  It may be
seen that the TPAGB tracks pass through the area where dust forming
atmopsheres exist on Figure~\ref{hrstars}.  The pulses may
allow for periodic dust formation by increasing the density around the
condensation point (Bl\"{o}cker 1995b) and carbon
stars evolving on these tracks can reach $T_{\mathit{eff}}<2500K$
(Marigo 1996), increasing the likelihood of nucleation.
The future evolution of the star may depend on where the thermal
pulses occur as they ascend the AGB. If the TP occurs just before the AGB
ends, the star will experience dredge up and may become C rich. A
superwind is then thought to occur with ${\dot M} \sim 10^{-4}
M_{\odot}yr^{-1}$ (Schonberner \& Steffan 2001; Bl\"{o}cker 1995b).
If the TP occurs after the
AGB ends (characterised by an increase in $T_{\mathit{eff}}$ and a decrease
in $\dot M$ (Fujii et al 2001)), then it will evolve on a constant luminosity
track for $\sim10^3$years to a Planetary Nebula stage (PN) - and the star
is now PAGB (Post AGB). This is characterised by an increasing $T_{\mathit{
eff}}$ and a decreased mass loss with
temperatures reaching above 30,000K.  The central star begins to
ionize the surrounding material to create a PPN (Proto-Planetary
Nebula).  This phase does not seem suitable for dust formation.
It does not satisfy the low temperatures or the high mass loss rates
needed, so we can ignore this as a significant phase of dust
production (although circumstellar dust shells are seen around PAGB
stars, they are thought to have survived from their previous AGB phase
(Lorenz-Martins et al 2001).  If the TP occurs when He burning
has ceased then
the star will evolve on a white dwarf cooling track and will again not be
suitable for dust conditions.

So in these later evolutionary stages we have a clear candidate for a
dust formation site - stars in the TPAGB and the superwind phases.
Those stars which have had their composition changed from O rich to C
rich are now more likely to form C rich dust.  It is interesting to note
that from Section~\ref{nuc}, not only is it difficult to reach the 
low temperatures in stars for silicates to form but we are now faced
with the problem that those O rich stars are likely to become C rich
in the most significant dust formation stages!  This would infer that
silicate dust may not be the dominant component of returned dust from
stars as previously thought (Bode 1988), but instead SNe may be
responsible for most of the silicate dust we see in the ISM. Dust
formation in the TPAGB phase will also be highly dependant on metallicity 
as dredge ups are more efficient at low Z, hence the TPAGB is reached
earlier on in the stars' lifetime and more C stars are
formed (Dwek98; Groenewegen 1999; Marigo 1998; Marigo et al 1999).  
The enhanced conversion of O stars to C stars at low
Z may then imply a large 
abundance of carbon dust in the early universe (although at Z=0.001 we
need at least $1\times 10^8$yrs for $M \leq 5M_{\odot}$ to reach the
TPAGB phase).

It is worth mentioning that if it is mainly carbon dust forming in
stars during the TPAGB phase then it may be formed from freshly
synthesised elements in the star and may be considered differently to
dust formed from pre-existing heavy elements in stellar atmospheres.
This may have consequences in models which separate the formation of
dust into two different modes.  From E01, we can define a dust formation 
efficiency (i.e. how much of the available heavy elements will be
incorporated into dust) depending on whether those heavy elements are from
freshly synthesised (e.g. in SNe) or pre-exisitng material (e.g. in
stellar winds).  We label these dust condensation efficiencies
$\chi_1$ and $\chi_2$ respectively.
The TPAGB phase may thus be
considered as a $\chi_1$ source instead of a $\chi_2$ as although the
dust is being formed in a stellar wind, its source of heavy elements
is dredged up from freshly synthesised carbon from the core.  Although 
the ejecta is likely to be mixed, the carbon dredged up during the
TP-AGB phase is a way of producing fresh elements at low metallicity,
i.e. the production of both silicate and carbon dust may depend on
metallicity.  The
implications of this is investigated in Section~\ref{sec:model}.

\subsection{Estimating the Condensation Efficiency in Winds}
\label{sec:chi2}

\begin{table*}
\centering
\begin{minipage}{14cm}
\begin{tabular}{ccccccccccccc} \hline
$M_i$ &$\tau_m$ & $\Delta M_{ej}$ & $\Delta M_{SW}$ & $mp_C$ & $mp_Z$
& $M_{dust}$$^a$ & $M_{dust}$$^b$ & Total $M_{dust}$ & Implied $\chi_2$
\\ \hline
$1$ & $10^4$ & 0.5& $0.5$ & 2.4E-5 & 2.4E-5 & 8.0E-6 & 5.2E-6 & 1.3E-5 & 0.55\\ 
$3$ & $350$ & 2.4 & 2.4 & 3.5E-2 & 3.0E-2 & 6.0E-4 & 1.9E-2 & 2.0E-2 & 0.65\\
$4$ & $160$ & 3.3 & 3.3 & 1.5E-2 & 2.0E-2 & 2.3E-4 & 1.5E-2 & 1.5E-2 & 0.76\\ 
$5$ & $94$ & 4.2 & 4.2 & 5.0E-3 & 1.2E-2 & 1.7E-4 & 9.0E-3 & 9.2E-3 & 0.76\\ 
$9$ & $26$ & 7.8 & 2.8 & 2.7E-2 & 1.7E-1 & 1.0E-4 & $<$6.0E-4 & $<$7.0E-4 & $<0.7$\\ 
$12$ & $16$ & 10.6 & 2.6 & 7.1E-2 & 6.9E-1 & 1.4E-4 & $<$6.6E-4 & $<$8.0E-4 & $<0.8$\\ 
$15$ & $11$ & 13.3 & 3.3 & 1.4E-1 & 1.32 & 2.6E-4 & $<$3.4E-4 & $<$4.8E-4 & $<0.6$\\
$20$ & $8$ & 17.8& 5.0 & 2.2E-1 & 2.73 & 5.6E-4 & $<$2.4E-4 & $<$8.0E-4 & $<0.8$\\ 
$25$ & $7$ & 22.3 & 9.4 & 3.0E-1 & 4.48 & / & / & / & /\\ 
$40$ & $4$ & 38.2 & 30.0 & 4.88 & 8.01 & / & / & / & /\\ \hline
\end{tabular}
\end{minipage}
\caption{\label{tab2}\small{The maximum mass $M_{dust}$ lost in winds from stars,
$M_{\odot}$, which satisfy the conditions for dust nucleation,
for $Z=Z_\odot$.  The columns are (1) Initial mass.  (2) Main sequence lifetime in Myrs
(Schaerer et al, 1993).  (3) Total ejecta mass lost including SW's and
SNe (Maeder, 1992).  (4) Total mass lost in SW's.  (5) Total carbon yield
for low mass stars
taken from Marigo (2000) and for high mass stars from Maeder (1992).
(6) As (5) for heavy elements, Z.  (7) Mass of dust from winds in the RG/E-AGB
phase. Mass loss rates and effective temperatures
used to estimate the maximum dust mass that could form were 
taken from the models in Schaller et al (1992).  (8) Maximum dust mass 
in the TP-AGB/superwind phase.  Carbon dust formed in TPAGB
assumes $\sim 80\%$ of the freshly synthesised carbon satisfies the nucleation
conditions.  (9) Total dust 
mass produced.  (10) Implied condensation efficiency calculated from
Equation~\ref{fraction1}.}}
\end{table*}

To estimate the condensation efficiency in winds we can use the models 
in Schaller et al (1992) and Marigo et al (1996) to give the mass fraction lost in
winds (as listed in Table~\ref{tab2}), and the fraction which satisfies the
conditions for
nucleation given in Section~\ref{nuc}. 
Equation~\ref{fraction1} will then give an estimate of $\chi_2$
(although this is actually a maximum as we assume $100\%$ condensation
wherever the nucleation conditions are satisfied).
\begin{equation}
\chi_2={\mbox{Mass into dust in stars} \over \mbox{Mass
of heavy elements lost in Stellar Winds}}
\label{fraction1}
\end{equation}

It must be emphasised that a more detailed definition of the dust 
condensation efficiency in stellar winds would be 
to account for the free carbon in the ejecta (i.e. C not locked up in
CO) and any available Si rather than just take the mass of heavy
elements ejected as was done here.  This has been done in a much more
detailed manner in Dwek98 but in high redshift galaxies, it would 
be difficult to determine the composition from the sub-mm fluxes
only.  Thus we try and estimate the maximum dust mass which could be
created in winds and ignore detailed composition calculations.  The new
dust formed in the TP-AGB phase is likely to be carbon based whereas
in the superwind phase, the dust is likely to be silicate based
depending on the C/O ratio in the star.  The majority of any
pre-exisiting dust in the ejecta is likely to be silicates.

To determine the average condensation efficiency over a stellar
population we need to sum over the stellar mass spectrum, with
Salpeter initial mass function, $\phi(M) \propto M^{-2.35}$.
The uncertainty of whether or not the TPAGB phase represents $\chi_2$ or
$\chi_1$ does not allow a clear estimate to be made, and initially we
estimate $\chi_2$ for the models in Schaller et al (1992), and
$\chi_2^*$ which also includes the TPAGB and the superwind phase.  

So for solar
metallicity and using Table~\ref{tab2}, 
\[
\chi_2 \sim 0.16
\]
\[
\chi_2^* \sim 0.45
\]
If the condensation efficiency of nucleation $<100\%$ then this would
decrease $\chi$.  Published
estimates are $\sim 0.3-0.7$ (Dominik et al, 1993) for condensation of Si, Fe,
Mg, O and $\sim 0.3$ for carbon (Cadwell et al, 1994) in stellar
atmospheres.   
The numbers given in Table~\ref{tab2} are a maximum and
could be as low as 0.1 if we assume only 30 per cent nucleation efficiency
when conditions are right.  This is a minimum limit for an elementary
model (E01) to reach the observed dust masses in the solar
neighbourhood.  From these simple arguements, it appears that the
TPAGB and superwind phases may be the most significant dust production
phase in the stars lifetime. 

\subsection{Estimating the Condensation Efficiency in SNe}
\label{sec:chi1}
Recent work on the nucleation of dust grains in supernova gas has
begun to shed some light in this area (Kozasa et al, 1997; Clayton et al,
2001; TF01).  Using SN1987A observations as
a basis, they all predict rapid growth of dust grains around 500-600
days after the initial explosion (as shown in SN1987A by the increase in the
luminosity emitted in the Infrared at these times).  The observational 
evidence for dust condensation in SN1987A is promising (for a detailed 
discussion see Wooden, 1997) but it nevertheless has a featureless
emission spectra with no Si, C or O attributes.  This is thought to be 
accounted for by iron type grains which is also suggested by the
similarities in the iron and dust filling factors of the ejecta
(Wooden, 1997).  The total mass of dust observed in SN1987A from IR
emission is $\sim 10^{-4}M_{\odot}$ although the elemental depletion
seen in the ejecta suggests this should be a lot higher (Dwek, 1992).
This discreprancy can be explained if a colder population of dust
grains exists and survives in ejecta clumps (Lucy et al, 1991).
These grains would radiate at longer wavelengths and could produce a
further $M_{\odot}$ of dust not detectable in IR.  Future Sub-mm
observations of SNR's could detect this cooler
dust and if discovered may suggest condensation efficiencies of
$\sim 100$
per cent.  There is other observational evidence to
support this view with dust emission seen in the young SNR Cassiopiea A.
IR observations of Cass A indicate that $10^{-4}M_{\odot}$ of
dust is present (Arendt et al, 1999; Douvion et al, 2001a,b) which is
much less than expected from theory.  Two
'recently' observed supernovae SN1993J and SN1994I
show no sign of dust condensation in their ejecta and unlike SN1987A,
which began its SN phase as a blue supergiant, 
these progenitor stars were red supergiants (Wooden, 1997).  It
remains to be seen if SN1987A and CassA are special cases.

Theory is still a long way from reliably estimating the condensation
efficiency of dust in SNe of different progenitor mass and
evolutionary phases.  The most acceptable theories begin with
nucleation via chemical or kinetic rates, but still can not exactly match the
observations of SN1987A (with that of Kozasa et al
(1989, 1997) predicting lower gas temperatures than actually seen).
Interestingly the models in Clayton et al (2001) allow for a greater
production of carbon dust even in ejecta with $C/O<1$ compared to the
more classical nucleation models (Kozasa et al, 1989, 1997).  Again
there may be the possibility of more carbon dust produced than
silicates.
We take the elementary model published in TF01 to estimate the condensation
efficiency in SNe (although this is based on the nucleation theory
given in Kozasa et
al, which has been shown to be incorrect for SN1987A).  The
model gives the dust mass produced
per supernova as a function of the mass of the progenitor star and the metallicity.
TF01 have modelled two cases of dust production - (a) and
(b) - corresponding to low and high energy SNe explosions respectively.  

To estimate $\chi_1$ for these two cases,
we require information on the heavy element
yield from the ejecta (as we are interested in finding the condensation
efficiency of dust forming from freshly synthesised
elements, Equation~\ref{fraction2}). 
\begin{equation}
\chi_1={\mbox{Mass of dust formed in ejecta} \over \mbox{Mass of
freshly formed heavy elements in ejecta}}
\label{fraction2}
\end{equation}
The yields were obtained from the tables given in Woosley \& Weaver
(1995), for a grid of
stellar masses and metallicities $Z/Z_{\odot}=0, 10^{-4}, 10^{-2}, 1$.
Weighting the values over the stellar mass spectrum for solar
metallicity, gives: 
\[
\chi_1(a) \sim 0.22
\]
\[
\chi_1(b) \sim 0.23
\]

Other theoretical models suggest efficiencies of $\sim 0.1 - 0.3$
(Kozasa et al, 1991; Clayton et al, 2001) 
whereas observations
suggest $<0.1$ (Douvion et al, 2001b).
The available data allows
for a comparison of $\chi_1$ values at different metallicities as
discussed in the next Section, although we must stress that there are
no observations which clearly confirm the extent of any condensation
of dust in SNe (other than SN1987A and CassA), this is purely based on
theoretical models.

\subsection{The Dependence of Dust Sources on Metallicity}
\label{varychi}

Now we can ask the question does the condensation efficiency as
estimated here
vary with the metallicity evolution of a galaxy?  It seems obvious 
that there will be less dust formation at low Z due to higher
temperatures in the stellar atmopsheres but can we actually quantify
this relationship?  

It is useful at this stage to re-define the condensation
efficiencies used in previous Sections:
\begin{itemize}
\item $\chi_1$ - efficiency of dust condensation from new
heavy elements for the (a) high energy (b) low energy SNe explosions.
\item $\chi_2$ - efficiency of dust condensation from
pre-existing heavy elements in stellar winds of stars during their
RG/AGB phases.
\item $\chi_1^*$ - condensation efficiency of dust from stellar winds
of evolved stars during their TPAGB phase if we assume
the available heavy elements are freshly synthesised in the star.
\item $\chi_2^*$ - condensation efficiency of dust from pre-existing
heavy elements in the stellar winds of evolved stars during their RG/AGB phase
{\it and} their superwind/TPAGB phases (i.e. $\chi_2 + \chi_1^*$).
\end{itemize}
  
Using the same arguments as given in Section~\ref{sec:chi2}, the
condensation efficiencies in stellar winds, $\chi_2$ and $\chi_2^*$ at different
metallicities have been estimated.  The criteria for nucleation is 
the same with the mass loss rates and effective temperatures 
taken from Schaller et al (1992), Schaerer et al (1993), and Schaller et al
(1993).  The 
published models allowed estimates to be made for dust nucleation for
$Z=0.001, 0.008$ and $0.04$\footnote[5]{as distributed by the astronomical data
center at NASA Goddard Space Flight Center}.  Their corresponding
$\chi_2$ values are listed below in Table~\ref{tab3}.  $\chi_1^*$
represents the condensation efficiency in the TPAGB phase
{\em if} it is considered as a $\chi_1$ source i.e. freshly synthesised 
dust production.  $\chi_2^*$ groups the TPAGB into the stellar wind mode.

For $\chi_1$, the same process of estimating the condensation
efficiency seen in 
Section~\ref{sec:chi1} was applied to the yields at other metallicities
from Woosley \& Weaver (1995).  The values are given in
Table~\ref{tab3} where (a)
and (b) still represent the low and high energy cases.

For zero metallicity, it has been shown that the TPAGB phase does occur
but only at $T_{\em eff}>4500K$ (Marigo et al 1999).  Therefore the TPAGB
phase at low
metallicities is not considered here as a source of dust production.  For
$Z=0.008$, the star does appear to reach the low temperatures but for
a shorter period of time as compared to those at solar metallicity (as 
shown in Marigo et al (1996)).  The problems in trying to calculate the
condensation efficiency at different $Z$ become apparent when trying
to determine the mass loss rates and the time spent in the nucleation
conditions.  With lack of any concrete information it is assumed that
the TPAGB properties at $Z=0.001, 0.008, 0.04$ are similar to those at
$Z=0.02$ but at lower metallicities less dust is formed per star due to less
time being spent in this phase.  Note that this will be a maximum
estimate of $\chi$.
There is no significant correlation between the metallicity and the
condensation efficiency in SNe so we assume $\chi_1$ is
roughly constant, $\sim 0.23$.  The approximate variation of $\chi_2$, 
$\chi_2^*$ and $\chi_1^*$ with metallicity is given below,
\[
\chi_2(Z)=-162Z^2 +9Z
\]
\[
\chi_2^*(Z)=-676Z^2+33Z
\]
\[
\chi_1^*(Z)=-5807Z^3-612Z^2+21Z
\]
\begin{table}
\centering
\caption{\label{tab3}\small{The maximum stellar weighted condensation
efficiencies, $\chi_2$, $\chi_2^*$, $\chi_1^*$ and $\chi_1$ in stars at
different metallicities, $Z/Z_{\odot}$.}}
\begin{tabular}{cccccccc}\hline
$Z/Z_{\odot}$&$0.0$&$10^{-4}$&$10^{-2}$&$0.05$&$0.2$& $1.0$&$2.0$\\ \hline
$\chi_2$& 0 & / & / & / & 0 & 0.16 & 0.10\\
$\chi_2^*$&0&/&/&0.10&0.10&0.45&0.21\\
$\chi_1^*$&0&/&/&0.10&0.11&0.22&0.22\\
$\chi_1(a)$&0.46&0.15&0.18&/&/&0.22&/\\
$\chi_1(b)$&0.27&0.18&0.21&/&/&0.23&/\\ \hline
\end{tabular}
\end{table}

\section{Modelling Chemical Evolution}
\label{sec:model}

Using the notation given in Pagel (1997) and Henry et al
(2000), we can follow the buildup of
heavy elements over time in a closed galactic system.  Initially we
have a mass of gas, 
g, which is converted into stars, s, via a star formation rate - SFR, 
$\psi(t)$ and an IMF, $\phi(m)$.  The total mass of the closed system
will then be 
\[
M=g+s
\]
Using an infall model (such as the one described in Dwek98), the
extra dilution of the ISM limits the metallicity reached and hence the 
dust mass will be slightly less.  Edmunds \& Eales (1998) show
that the simple model produces the maximum
dust yields for a given gas fraction hence we assume no inflows or
outflows (the effects of which are discussed elsewhere (e.g. Koppen
and Edmunds, 1999; E01)).
The gas mass, g will be depleted by
star formation and enriched by the mass ejected from stars, e(t).
\begin{equation}
{dg \over{dt}}=-\psi(t)+e(t)
\label{ginc}
\end{equation}
The evolution of heavy elements in the ISM ($Zg$) is governed by the
ejection of metals by stars ($e_z$) and those going into forming stars.
\begin{table}
\centering
\caption{\label{yields}\small{The integrated yields, $p_x$ of heavy
elements ($p_Z$) and carbon ($p_c$) from massive and intermediate mass
stars at different metallicities, Z.}}
\begin{minipage}{8.5cm}
\centering
\begin{tabular}{cccc} \hline
Mass Ranges& Z & $p_Z$ & $p_c$ \\ \hline
$1-5M_{\odot}$\footnote[1]{From Marigo, 2000}& 0.004& $3\times10^{-3}$ &
$2\times10^{-3}$ \\	
&0.02& $5\times10^{-4}$&$4\times10^{-4}$ \\
$9-120M_{\odot}\footnote[2]{From Meader,
1992}$&0.001&$0.018$&$2\times10^{-3}$ \\
&0.02&0.015&$8\times10^{-3}$\\ \hline
\end{tabular}
\end{minipage}
\end{table}
\begin{equation}
{d(Zg) \over{dt}}=-Z(t)\psi(t)+e_z(t)
\end{equation}
Therefore the metallicity evolution is described by
Equation~\ref{zinc} 
\begin{equation}
{dZ\over dt}={e_z(t)-Z(t)e(t)\over g}
\label{zinc}
\end{equation}
Assuming that mass loss occurs suddenly at the end of
stellar evolution, the ejected mass and the ejected heavy elements are
(Tinsely 1980)
\begin{equation}
e(t)=\int_{m_{\tau_m}}^{m_2}{\left[m-m_{R}(m)\right]\psi(t-\tau_m)\phi(m) dm}
\label{chemg}
\end{equation}
\begin{eqnarray}
e_z(t)&=& \int_{m_{\tau_m}}^{m_2}\bigl({\left[m-m_{R}(m)\right]
Z(t-\tau_m)+mp_z}\bigr) \nonumber \\
      & &\mbox{}\times~\psi(t-\tau_m)\phi(m)dm
\label{chemz}
\end{eqnarray}
where $mp_z$ is the stellar yield of heavy elements from a star of initial
mass m (values taken from Maeder (1992) and Marigo (2000) for massive
stars and IMS respectively).  $m_R(m)$ is the remnant mass,
$\tau_m$ is the lifetime of a star of mass m such that a star formed
at $t-\tau_m$ has died at time t, and $m_{\tau_m}$ is the stellar mass
whose $\tau_m$ corresponds to the age of the system.  Thus we have
taken into account the finite age of the stars ejecting the material
into the ISM.

The second term in Equation~\ref{chemz} gives the integrated yield, 
$p_x$ which defines the mass fraction of stars formed in the mass
range $m_1$ and $m_2$ and eventually expelled as new element $x$,
Equation~\ref{Px}, Table~\ref{yields}.
\begin{equation}
p_x=\int_{m_1}^{m_2}{mp_x(m)\phi(m)dm}
\label{Px}
\end{equation}
$\phi(m)$ is the initial mass function which is the normalised
Salpeter IMF with b=1.35
\begin{equation}
\phi(m)=\left(1-b \over{m_2^{-(1-b)}-m_1^{-(1-b)}}\right)m^{-(1+b)}
\end{equation}
where $m_2=120M_{\odot}$ and $m_1=0.1M_{\odot}$.  It is not straightforward to
solve these equations without making some simple assumptions so 
we approximate the remnant mass $m_{R}(m)$ from Talbot et al (1973)
such that $m_R(m)=0.11m+0.46$ for $m \le 6.86 M_{\odot}$ and
$m_R(m)=1.5 M_{\odot}$ for $6.86 < m\le 120M_{\odot}$.
The assumed star formation rate, $\psi(t)$ scales as the gas density
to some power $\eta$ where
\begin{equation}
\psi(t)=kg^{\eta}
\end{equation}
and the constant k we label as the star formation efficiency. $\eta$
is thought to lie between 1 and 2 - although here we assume
$\eta=1$ for simplicity.

The code was written with a logarithmic timestep with initial
conditions for the gas and metallicity of the system:  
\[
g(t=0)=M_{tot}=1~~;~~Z(t=0)=0
\]
The
yields from Maeder (1992) and Marigo (2000) were interpolated to find
the stellar yield $mp_z$ corresponding to a given metallicity and
mass.  The stellar lifetimes $\tau_m$ were taken from the Schaller et
al papers and were also interpolated to find the relevant value for m
and Z.  Equations~\ref{ginc} \& ~\ref{zinc} were then solved at each
timestep with each increment of Z and g added to the previous
solution. 

\subsection{Dust Evolution}
\label{dust}

\begin{figure*}
\begin{minipage}{11cm}
\epsfxsize=11cm
\epsffile{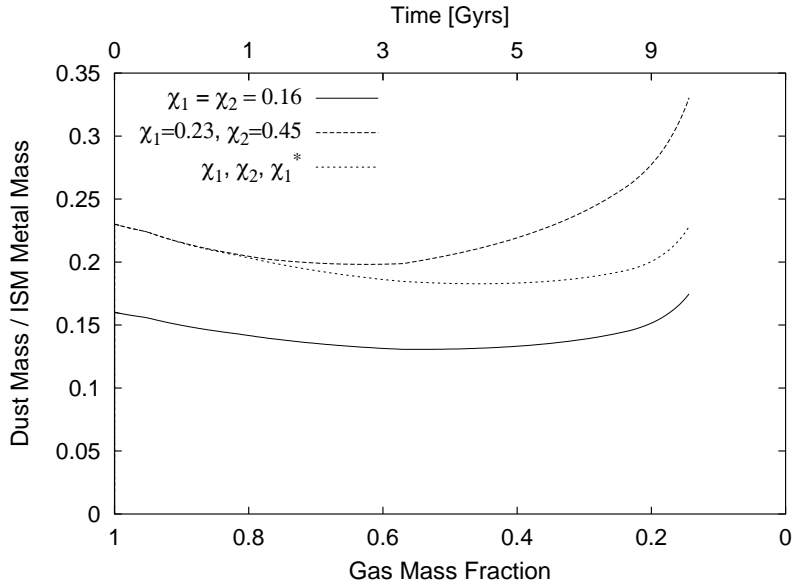}
\end{minipage}
\caption{\small{We compare the evolution of the ratio of grain core
mass to ISM
metal mass ($y/z$) vs gas mass for different dust
condensation
efficiencies in SNe and stellar winds.   This compares the
evolution of dust for a galaxy with equal contributions of dust
injected from stars and SNe ($\chi_1=\chi_2=0.16$) with a galaxy in
which this contribution is not equal by solving Equation~\ref{yg}.  The
dust production from
stellar winds becomes significant as the metallicity of the system
increases which is shown by the late increase in the $\chi_2$ model at 
$g \sim 0.2$.  We also plot the dust produced
during the TPAGB phase (using Equation~\ref{tpmodel}) with $\chi_1$,
$\chi_2$ and $\chi_1^*$ equal to 0.23, 0.16 and 0.22 respectively.  
In this model, the dust
produced during the TPAGB is considered as a separate dust mode - one
in which freshly synthesised carbon is produced and is now a
$\chi_1$-type source. The age of the system is shown on the z-axis
with initial condition t=0, g=1 for a closed box model with simple
linear in density SFR prescription giving g=exp(-$\alpha$ kt).}}
\label{comparemodel}
\end{figure*}
From E01, the elementary model uses the instantaneous recycling
approximation and follows the evolution of gas mass (g),
metallicity (Zg) and grain core mass (yg) as a mass ds forms
stars as in equations~\ref{gg},~\ref{zg} \&~\ref{yg}.
The destruction of grains is ignored here as is also mantle growth
onto the grains in the ISM (previously included in E01, Dwek98).
The terms in Equation~\ref{yg}
represent the relative
contributions of dust sources to the grain mass as a mass ds forms
stars.  The first term gives 
the contribution from SNe, the second from stars and the third from
grains lost in forming stars.
\begin{equation}
dg=-\alpha ds 
\label{gg}
\end{equation}
\begin{equation}
dZ={p\alpha ds\over g}
\label{zg}
\end{equation}
\begin{equation}
d(yg)=p^\prime \chi_1 ds+(1- \alpha) \chi_2 zds-yds
\label{yg}
\end{equation}

However if the TPAGB can be considered as a 'new' dust production
mode then this will affect the elementary model.  We can now propose an
additional term allowing for the grain cores formed in
the TPAGB phase with condensation efficiency, $\chi_1^*$ and
the fraction of C stars in a population, $\gamma$.  $\gamma$ will
depend strongly on the metallicity - the number of C stars increases
at lower Z due to an increase in the efficiency of
dredge-ups (Marigo et al 1999).  This
relationship is shown in Groenewegen (1999, Fig 3), and from this we can
roughly quantify a function for the variation of $\gamma$ with
Z, as in~Equation~\ref{gamma}. 
\begin{equation}
\gamma = \left[1 + \left({Z \over 0.0046}\right) ^{3 \over 2}\right]^{-1}
\label{gamma}
\end{equation}
This empirical relationship fits well the obsservations in which the
number of carbon stars in systems of different metallicity
(e.g. LMC/SMC, Fornax Dwarf Galaxy, etc.) are
identified by various spectroscopic methods. It is possible that the
relative number of carbon stars may also depend on the star formation
rate history (Gronenewegen, 1999) but detailed modelling would require 
excellent evolutionary tracks with reliable dredge up predictions.
Further work might usefully investigate such models but for our
present purposes this simple prescription should be adequate.

The integrated yield of carbon produced 
in low-intermediate mass stars during the TPAGB phase ($p^*$) can be estimated 
at different metallicities using Figure 12 in Marigo (2000).
Qualitatively, if we
assume that $10\%~(50\%)$ of the solar C/O value is carbon returned by 
these stars (Henry et al 2000) we obtain $p^* \sim 0.00073~(0.00365)$.  
The elementary model including the TPAGB phase is then
described by Equation~\ref{tpmodel}. 
\begin{eqnarray}d(yg*)&=& p\alpha\chi_1ds +(1- \alpha)(1-\gamma(Z))
\chi_2Zds - \nonumber \\
	&   &\mbox{} yds + \gamma(Z)(1-\alpha)p^*\chi_1^*~Zds
\label{tpmodel}
\end{eqnarray}

The solutions for the dust mass y(g) and y*(g) were
obtained numerically and are
shown in Figure~\ref{comparemodel}.  In this Figure we compare the
elementary model in E01 (Equation~\ref{yg}) to Equation~\ref{tpmodel}
by plotting the ratio of
the grain core mass (yg) to the metal mass (zg).  For the elementary
model case we use $\chi_1=\chi_2=0.16$ whereas for Equation~\ref{tpmodel},
the $\chi$ values are those defined in previous
Sections.  The increased dust
mass seen in Figure~\ref{comparemodel} initially, is due
to a higher $\chi_1$ value.  At
lower gas mass the increased dust mass is now due to a
higher $\chi_2$ (0.45).  This model therefore predicts more dust mass seen
throughout galactic evolution compared to the equal $\chi$ case of
the elementary model. 

We can include a time delay for the production of dust in stars
if we assume dust is all produced in stellar winds at the end of
stellar evolution (which is valid if our arguement in Section~\ref{nuc}
is correct).  This approach has been adopted before (Pagel, 1997;
Dwek98) and can be quite a useful approximation.  Using the lifetimes
of stars of different mass (Schaller et al, 1993) we can set the delay of dust
production, $\tau$ (in Gyrs) as:
\[
\tau_{d2}=0.1
\]
\[
\tau_{*}=2.4
\]
as the delay of dust production in stars producing $\chi_2$ and
$\chi_1^*$ dust respectively.  Thus the dust will evolve with time as 
\begin{eqnarray}
d(yg*)&=& p\alpha\chi_1\psi(t) + H(t-\tau_{d2})(1- \alpha)\nonumber \\
& &\mbox{} \times (1-\gamma(Z))\chi_2Z(t-\tau_{d2})~\psi(t-\tau_{d2})
\nonumber \\
	&   &\mbox{} - y\psi(t) + H(t-\tau_{*})\gamma(Z)
(1-\alpha)p^*\nonumber \\
& &\mbox{} \times \chi_1^*Z(t-\tau_{*})\psi(t-\tau_{*}) 
\label{tpdelay}
\end{eqnarray}
Equation~\ref{tpdelay} is not compared here as it is only important when
looking at temporal variations of dust evolution.  This is a useful
analytical approximation to follow the delay of dust production in low mass
stars but for a more detailed description we require further modelling 
(Section~\ref{sec:early}).  The TPAGB model is not significantly
different to the elementary model but will continue to be used here.
It may be useful in tracing the evolution of carbon and silicate dust
in the galaxy in future work.

\begin{figure*}
\epsfxsize=8cm
\subfigure[]{\epsffile{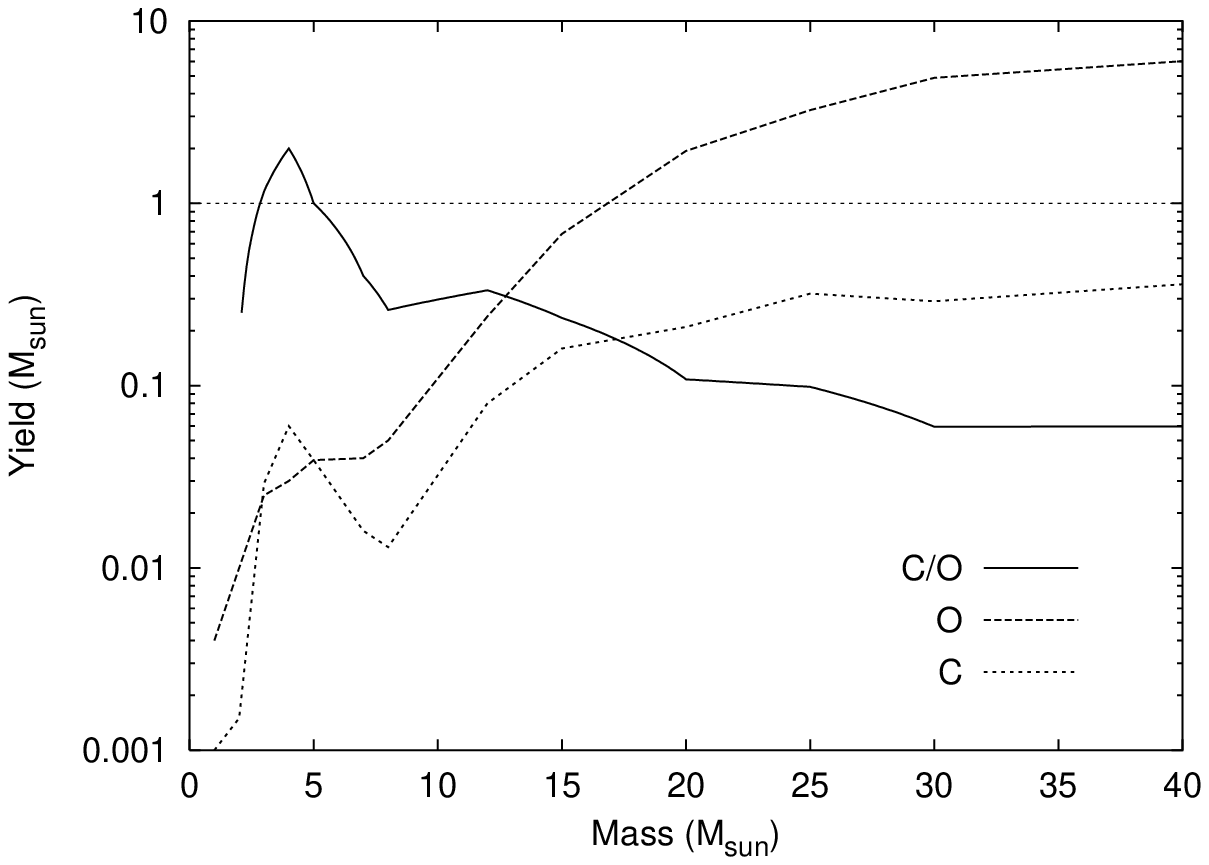}}
\epsfxsize=8cm
\subfigure[]{\epsffile{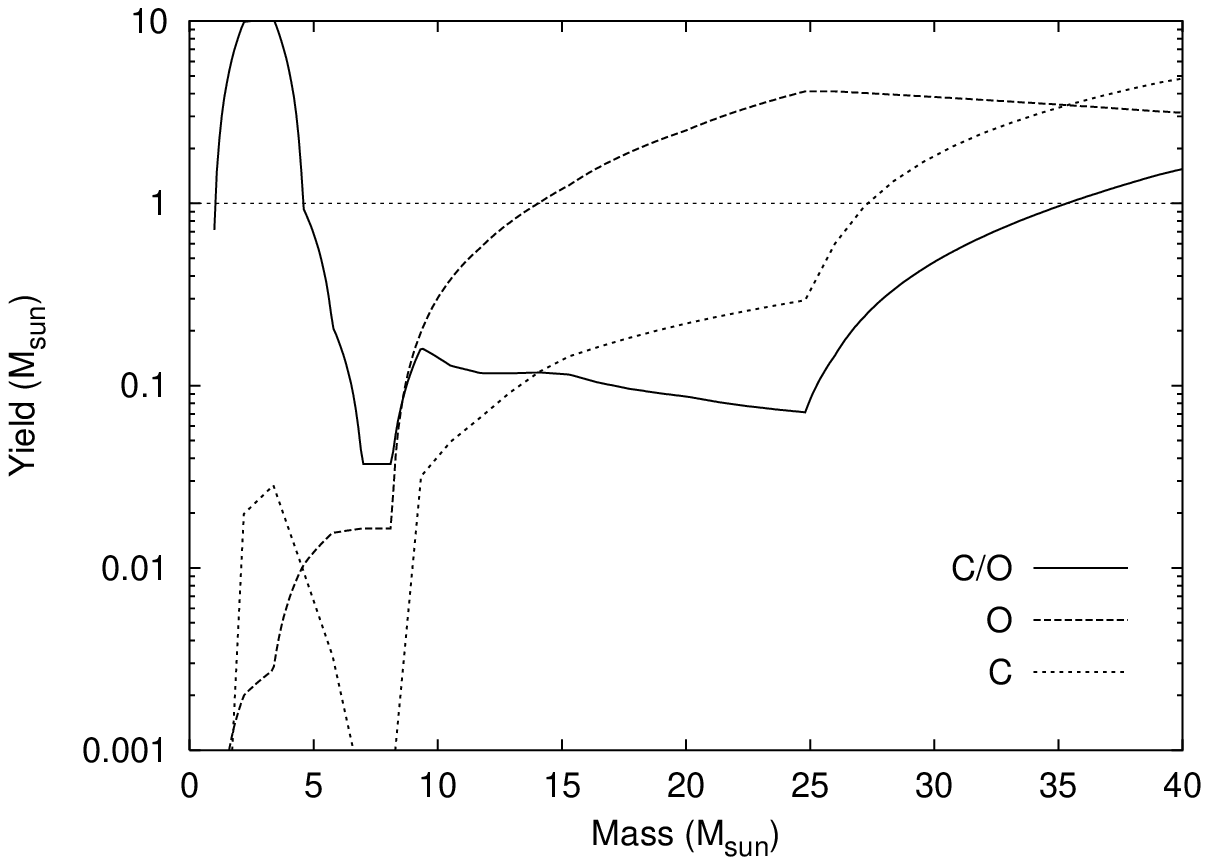}}
\caption{\small{The carbon and oxygen yields from stars with
$1 \le M \le 40M_{\odot}$ at $Z=Z_{\odot}$ from (a) Dwek98 and (b)
this work.  The C/O ratio is also shown in both figures.  This work
(b) shows a higher
C/O ratio for both low mass ($M<5M_{\odot}$) and massive stars
($M>25M_{\odot}$) than in the Dwek model.  The lack of heavies from
stars with $6<M<8M_{\odot}$ is much lower in this work than Dwek98 due 
to the negligible yields in AGB stars in this mass range (Marigo,
2000; van den Hoek \& Groenewegen, 1997).}}
\label{dweks6}
\end{figure*}
\begin{figure*}
\epsfxsize=8cm
\epsfxsize=8cm
\subfigure[]{\epsffile{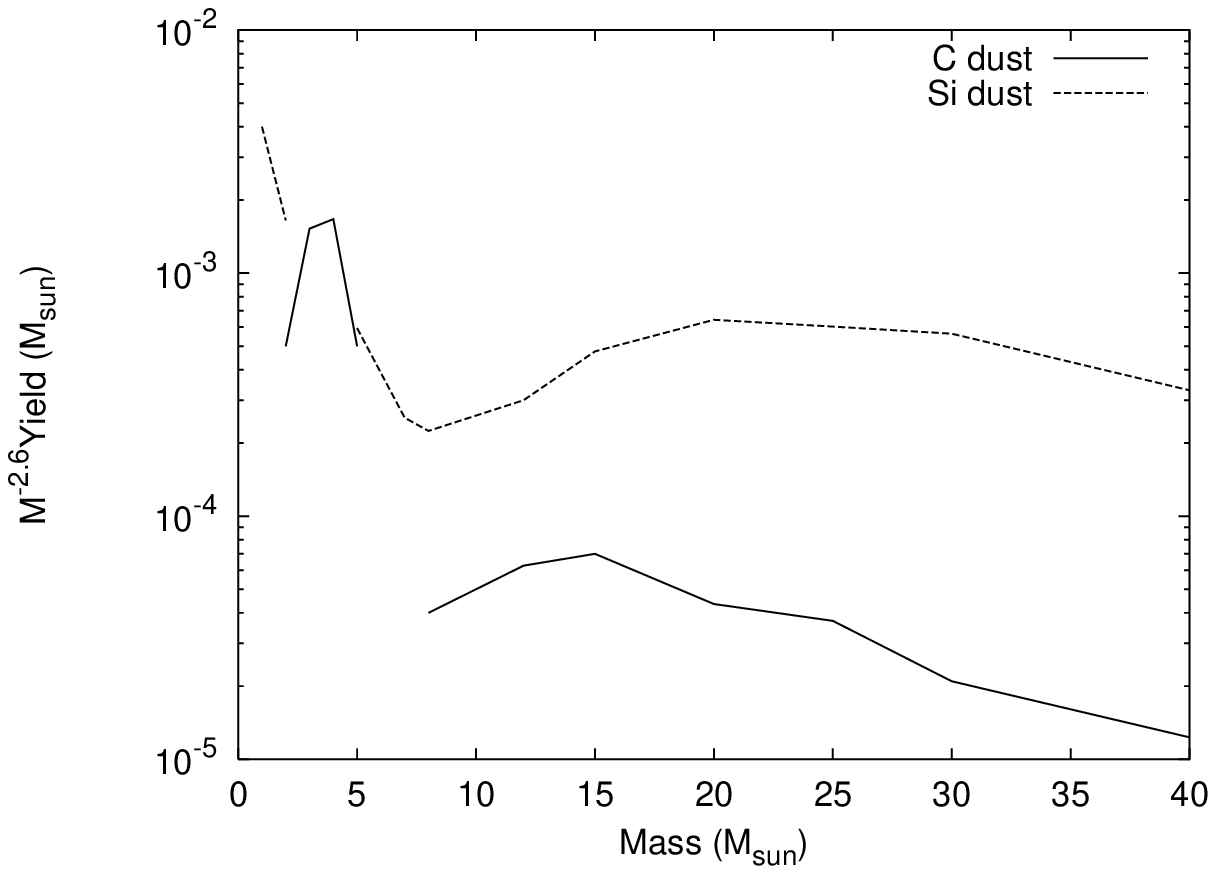}}
\epsfxsize=8cm
\subfigure[]{\epsffile{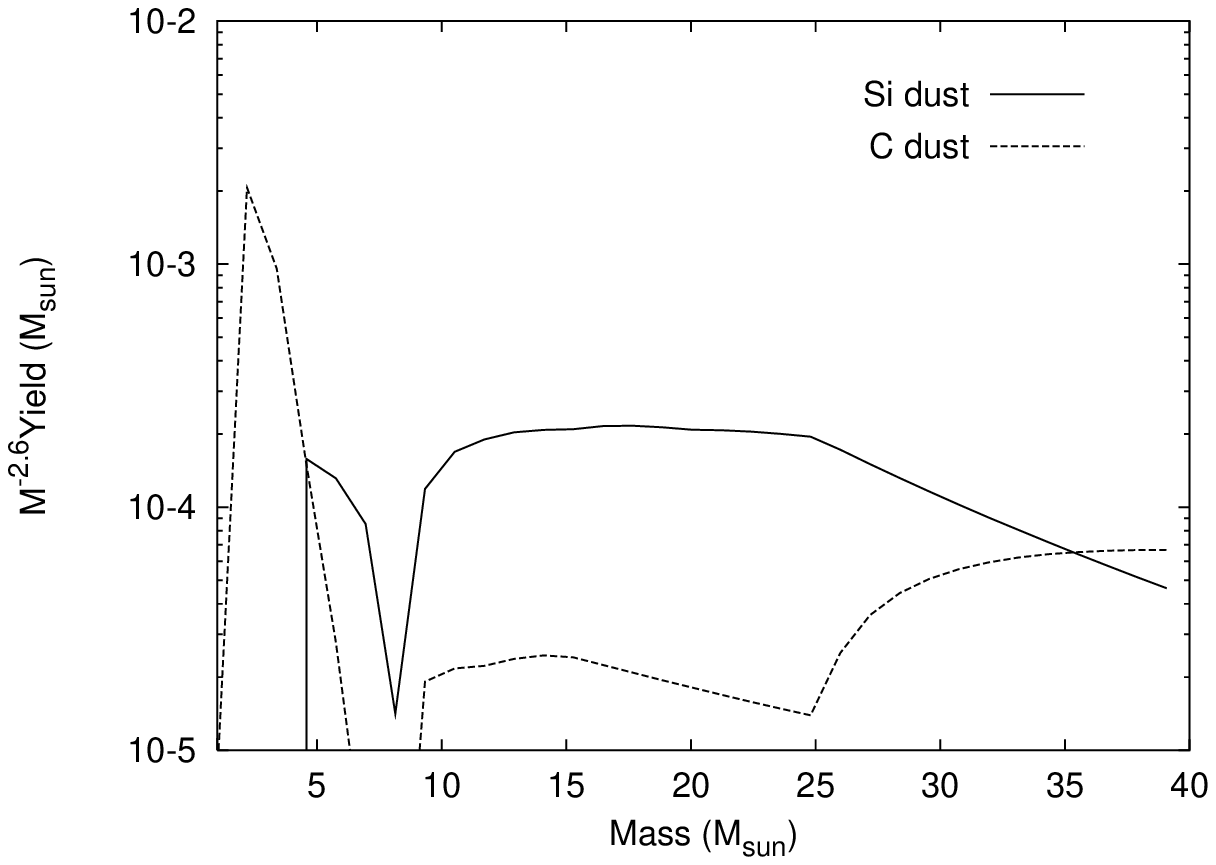}}
\caption{\small{Carbon and silicate dust yields weighted with a
stellar mass spectrum as in Dwek98 for $Z=Z_{\odot}$.  In (a) we
sketch the dust yields from Dwek98 Figure 6 and
plot our own results in (b) to compare the dust yields from the two models.  As in
Dwek98, the majority of carbon dust is produced by low mass stars
($<5M_{\odot}$) whereas most of the silicate dust is produced by
massive stars ($>10M_{\odot})$.  However, in this work, the amount of dust
produced by stars with $M>5M_{\odot}$ is less by a factor of $\sim 10$
than in Dwek98 due to differences in the yields and the condensation
efficiencies chosen.  This work uses the condensation efficiencies
listed in Table~\ref{tab2} for SW's and in Table~\ref{tab3} for SNe.
In Dwek98, the condensation efficiencies are $\sim 1$ in SW's and
$0.5-0.8$ in SNe.}}
\label{dweks7}
\end{figure*}

We can also compare our dust model with the work presented in Dwek98
which also follows dust formation via chemical evolution (although the model
presented here is a very simplified version with fewer parameters).  As 
both models assumes a fraction of the heavy elements will condense into 
dust, it is interesting to compare the differences in the yields used.
In Figure~\ref{dweks6} we plot the
carbon and oxygen yields used in both models from stars in the range
$1 \le M \le 40M_{\odot}$ for solar
metallicity.  In (a) we sketch Figure 6 in Dwek98 in
which the yields for IMS and massive stars are taken from Renzini \&
Voli (1981) and Woosley \& Weaver (1995) respectively, and in (b) we
plot the yields from this work as given in Table~\ref{tab2}.  This
work takes the more recent results for IMS yields from Marigo (2000)
in which a metallicity dependant mass loss prescription is used. This
produces higher carbon yields for IMS than those in Renzini \& Voli
(1981) and negligible carbon yields in stars with $M> 6M_{\odot}$ (as
is also shown in the TPAGB models presented in van de Hoek \&
Groenewegen (1995)).  This is thought to occur due to a drop in the TP
lifetime (hence lower number of dredge ups) due to HBB (Hot Bottom
Burning - see aforementioned papers for discussion).  The Renzini \&
Voli (1981) model also uses a lower mass loss parameter ($\eta=1/3$
instead of $\eta=4$ in Marigo (2000)), which will produce optimistic
yields for stars $M>3M_{\odot}$.  The differences between these
models are clearly seen in Fig~\ref{dweks6} (a) and (b) with this work 
having a high C/O
ratio produced in IMS but lower yields overall over the whole mass range.
This in turn leads to a lower dust yield in this work in comparison with
Dwek98 which is shown in Figure~\ref{dweks7}.  Again in (a) we sketch 
the model in Fig 7, Dwek98 in which the SMS weighted dust 
yields are shown for solar metallicity and compare with
this work in Figure~\ref{dweks7}(b).  As originally shown in Dwek98,
and reproduced in Figure~\ref{dweks7}(a) and (b), the majority 
of carbon dust is produced in IMS whereas the silicate dust is from
massive stars.  In (b) we have a greater yield of carbon dust from
more massive stars ($M>25M_{\odot}$) than (a) due to the higher C/O
ratio in massive stars in the Maeder(1992) yields.  It is clear that,
due to the particular stellar models and yields chosen, Dwek98 and
this work differ considerably in the prediction of dust production.
This probably indicates that a degree of caution should be used in
interpreting our overall results, but the underlying methodology
should be sound and detailed predictions will improve with better
stellar models.

\subsection{Dust Evolution with Metallicity}
\label{sec:compare2}

\begin{figure*}
\begin{minipage}{11cm}
\epsfxsize=11cm
\epsffile{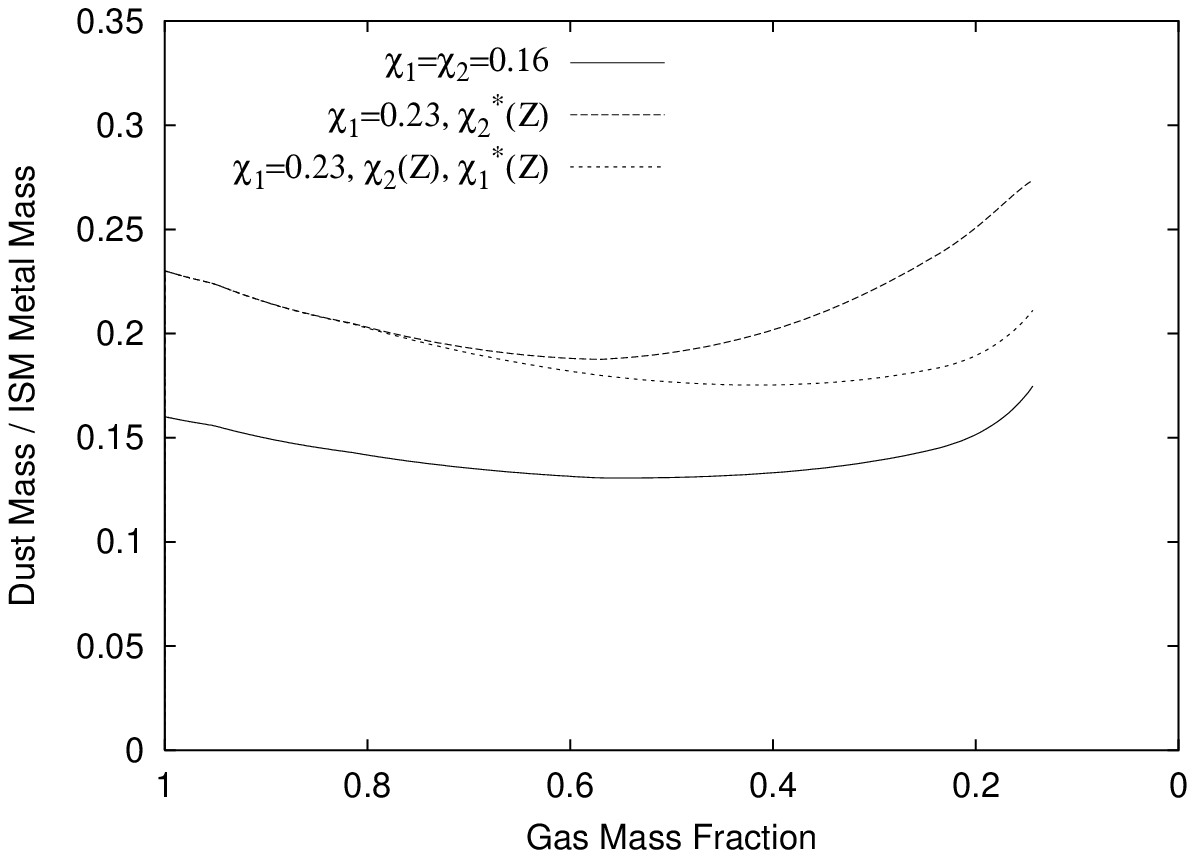}
\end{minipage}
\caption{\small{The comparison of dust evolution models with galactic
evolution.  Here we show
the grain core mass/ISM metal mass ($y/z$) for different 
dust condensation efficiencies which vary with metallicity as in
Table~\ref{tab3}. The relationships of dust production in stars and SNe
with metallicity are those 
given in Section~\ref{varychi} for $\chi_1$, $\chi_2$, $\chi_2^*$ and
$\chi_1^*$ and incorporating these into the model shows no significant 
difference to Figure~\ref{comparemodel} which uses constant values of $\chi$.}}
\label{compare2}
\end{figure*}
Putting the relationships from Section~\ref{varychi} into the
elementary model we obtain Figure~\ref{compare2}.
The variation of $\chi$
with metallicity as proposed here does not appear to differ
significantly from the
elementary model. The only differences
are in the increased grain core mass at low Z (due to a higher $\chi_1$) and at very
low gas masses (particularly for the $\chi_2^*$ relationship).  The
increased yield of carbon from the
models given in Marigo (1998) as compared to those in Henry et
al (2000) will also lead to an increased $\chi_1^*$ dust mass at low
g.  We conclude that the constant $\chi$
values derived in Sections~\ref{sec:chi2} and \ref{sec:chi1} are sufficient to
represent dust formation with metallicity. 

One of the main conclusions of the
elementary model is that the dust remains a fixed fraction of the metals
in the ISM (Edmunds \& Eales 1998) which
is also echoed here.  In Dwek (1980) and E01, this relationship is
proposed as $Z_{dust}=0.4Z_{ISM}$ which includes the growth of icy
mantles on top of the grain cores.  Without yet
considering mantle growth, the TP model gives 
$Z_{dust}=0.24Z_{ISM}$ which appears a little high.  For a solar
neighbourhood value gas mass of $\sim 0.15$,
we obtain a gas-to-dust ratio of $\sim 225$.  The observed value is
thought to be closer to 150 so this is well within observational
errors of a factor $\sim 2$.
It is expected that mantle growth onto the grain cores in the dense
ISM needs to be considered before a more realistic comparison can be
made. This leads to more complications as such a model will need
understanding of mantle
destruction and of the dense cloud fraction of the ISM in different
galaxies (E01).

\section{Dust Production In Early Galaxies}
\label{sec:early}
\begin{figure*}
\begin{minipage}{11cm}
\epsfxsize=11cm
\epsffile{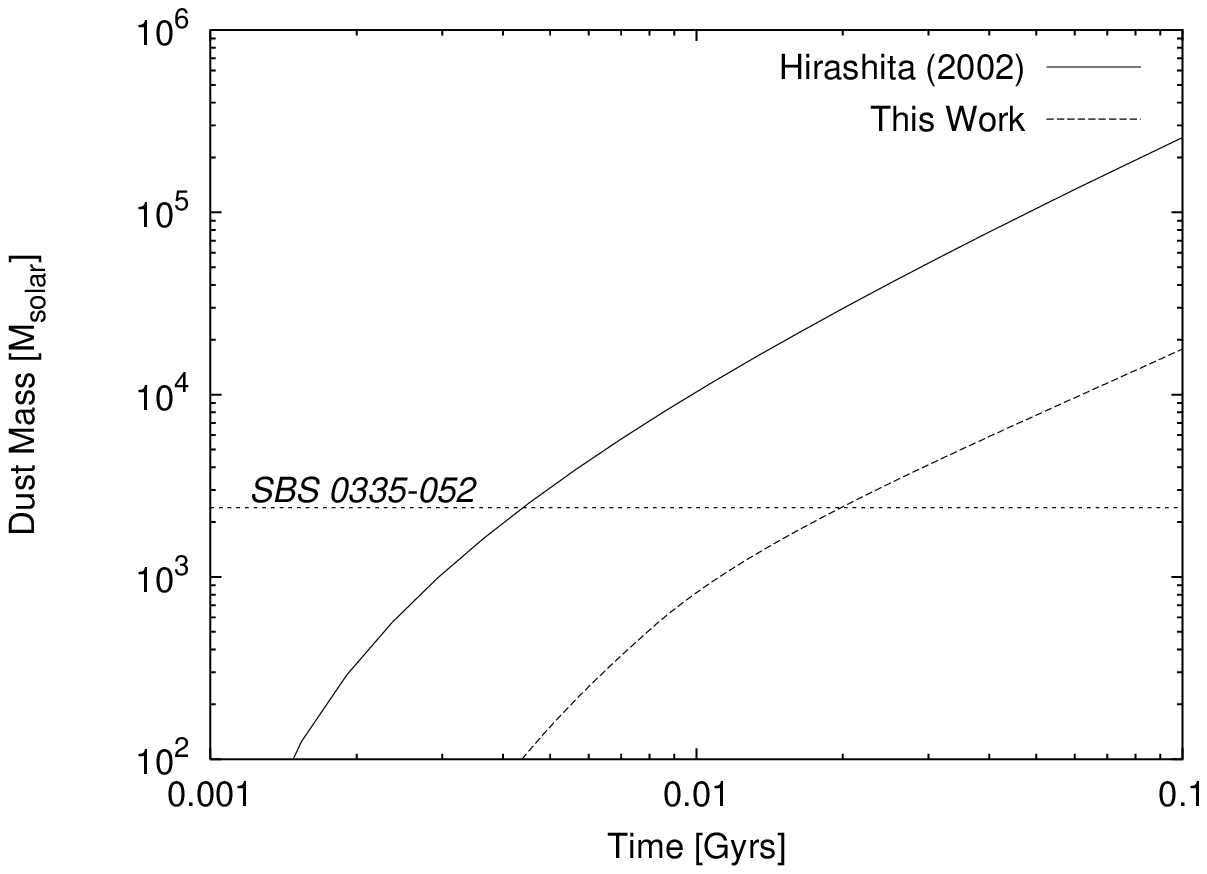}
\end{minipage}
\caption{\small{The evolution of dust mass with time for
a young starburst galaxy using our model with
$g=2\times10^{10}M_{\odot}$, $k=0.05$ (corresponding to a SFR of
$1M_{\odot}yr^{-1}$) and comparing to the model in Hirashita et al (2002).
The observed value of a galaxy (SBS {\small{0335-052}}) which matches
these properties is given also.}}
\label{hirash}
\end{figure*}
\begin{figure*}
\begin{minipage}{11cm}
\epsfxsize=11cm
\epsffile{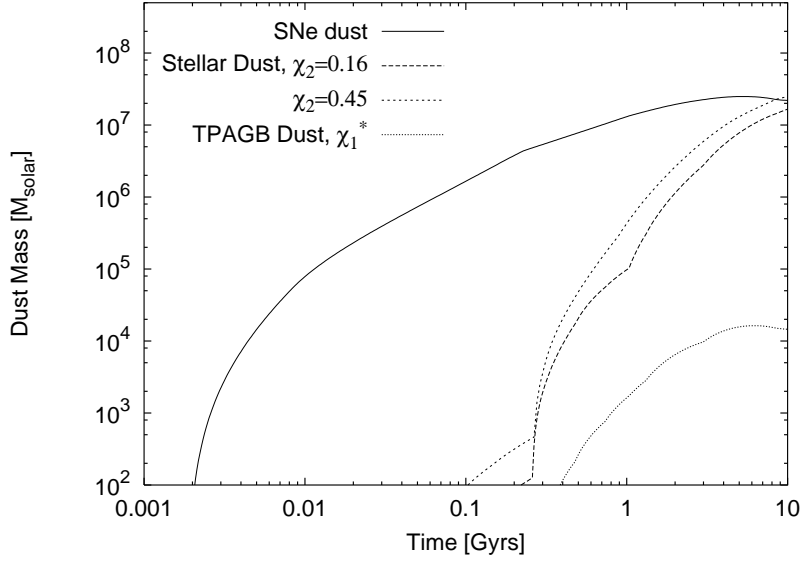}
\end{minipage}
\caption{\small{Evolution of dust mass (yg) with time for a
galaxy with initial gas mass $g=2\times 10^{10}M_{\odot}$ and SFR of
$5M_{\odot}yr^{-1}$ ($k=0.24$).  This galaxy is representative of the
Milky Way with gas mass fraction reaching 15 per cent after
$\sim11$Gyrs.  'SNe dust' corresponds to dust formed in supernovae only
with $\chi_1=0.23$, $\chi_2=0$.  Dust produced in stellar
winds with condensation efficiencies of $\sim 16$ and 45 per
cent are shown.  The TPAGB phase represents possible freshly
synthesised dust formed in intermediate mass stars only.}}
\label{delaychem}
\end{figure*}
\begin{figure*}
\epsfxsize=8cm
\subfigure[]{\epsffile{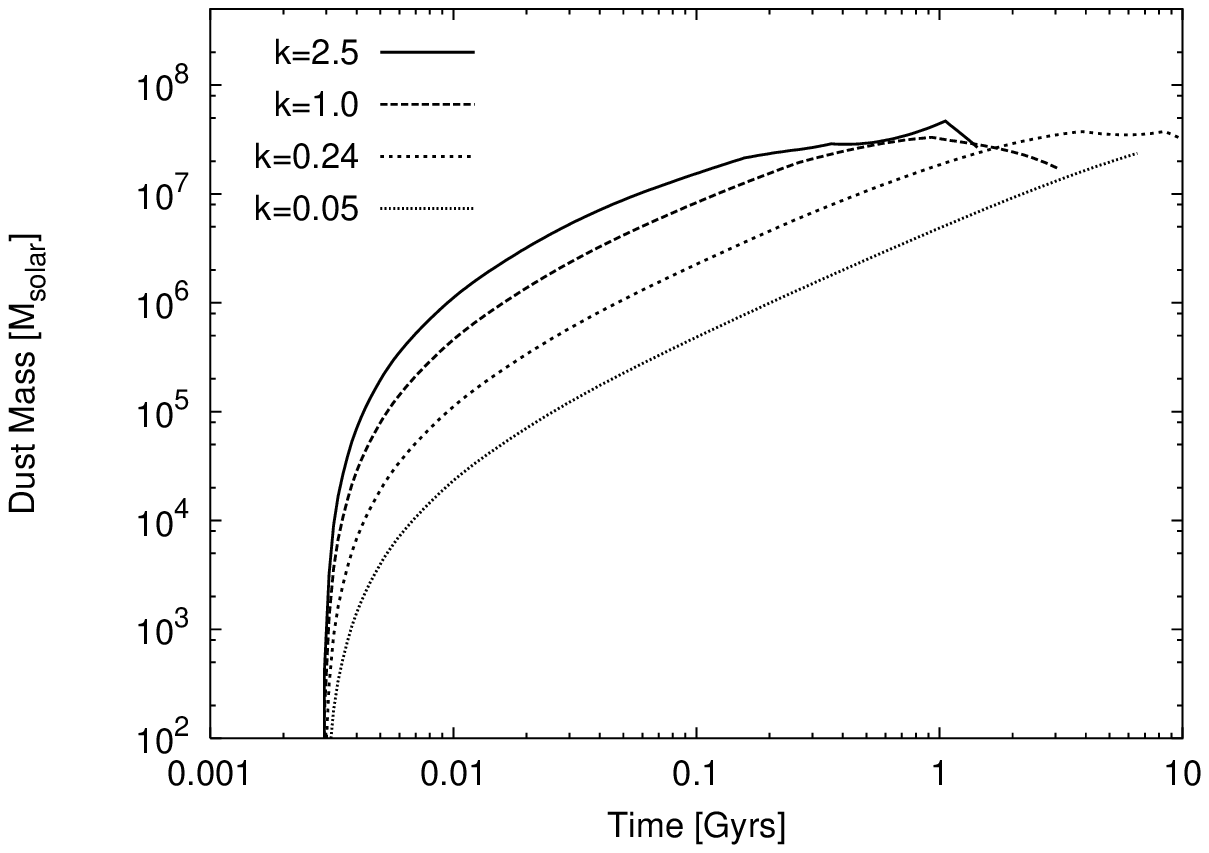}}
\epsfxsize=8cm
\subfigure[]{\epsffile{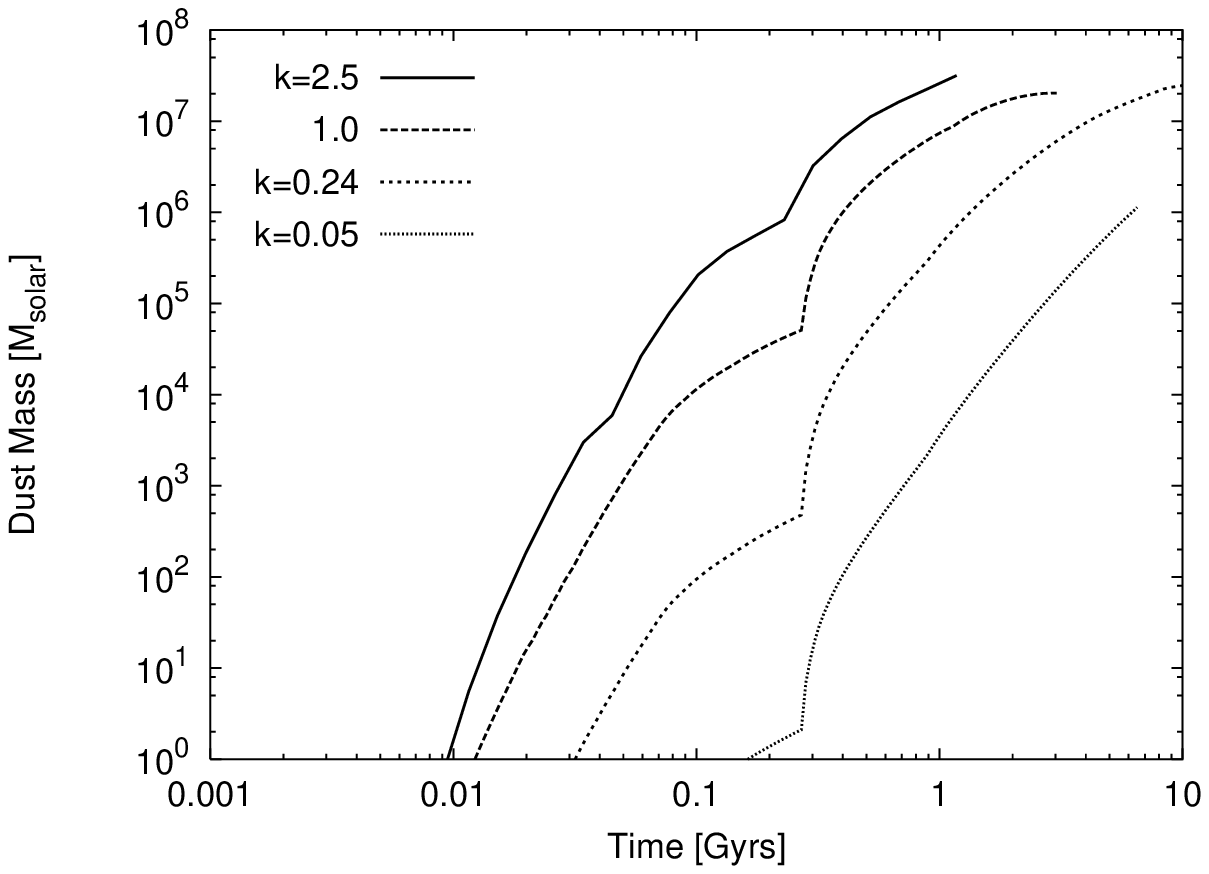}}
\caption{\small{The evolution of dust mass from (a) SNe
($\chi_1=0.23$) and (b) stellar
winds ($\chi_2=0.45$) with time for a given star
formation efficiency, k.  For initial
gas mass of $2\times10^{10}M_{\odot}$, we have a SFR for k=0.05, 0.24,
1.0, 2.5 of 1, 5, 20 \& 50$M_{\odot}yr^{-1}$ respectively.  This
sequence is likely to represent galaxies from spirals to
ellipiticals. }}
\label{k}
\end{figure*}
The earliest stages of star formation remain a mystery, but one
important factor is likely to be the role of dust.  The first heavy
elements could not
have been created until the first stars were formed, but how and when
did dust formation turn itself on?  Studying high redshift galaxies
and the obscuration effects of dust at these epochs should provide
some answers.  Some differences between early galaxies and our own are
quite obvious, 
these galaxies will certainly have higher temperatures and higher
supernova production rates than that found in the Milky Way.
Conditions at these earlier times, may have not been right for the
accretion of mantles within
the ISM with higher grain destruction rates.  The primary source of
dust cores here would be Type II
SNe, introducing heavy elements into
the ISM on timescales of $\sim 10^{6}$ yrs (if SNe are important as
dust formation sites).  

When modelling early galaxies we can no longer rely on the
instantaneous recycling assumption as used in Section~\ref{dust} to
describe the dust production in stars.  To 
be thorough, we require the detailed modelling of dust from stars
using yield calculations and the variation with metallicity of the
condensation efficiency in stars (Section~\ref{varychi}).  From 
Section~\ref{sec:model}, we can model dust in early galaxies using 
Equation~\ref{chemdust} where $\chi_2$, $\chi_1$ and $\chi_*$ are now the
$\it{unnormalised}$ condensation efficiencies.   The first term
corresponds to the dust
produced from recycled yields in stellar winds, the second the dust 
from freshly synthesised elements produced in massive stars and the
third from freshly synthesised carbon in IMS stars with $C/O>1$.
Equation~\ref{chemdust} was solved numerically with the code described in
Section~\ref{sec:model}.

\begin{eqnarray}
{dy \over
dt}&=&\int_{m_{\tau_m}}^{m_2}\bigl((1-\gamma(z))\left[m-m_{R}(m)\right] 
z(t-\tau_m)\chi_2\nonumber \\
            & &\mbox{} +mp_z\chi_1+\gamma(z){\left[m-m_{R}(m)\right]z(t-\tau_m)\chi^*mp_*}\bigr)\nonumber \\
     & &\mbox{}\times~\psi(t-\tau_m)\phi(m)dm -y\psi(t)
\label{chemdust}
\end{eqnarray}
\begin{figure*}
\epsfxsize=8cm
\subfigure[]{\epsffile{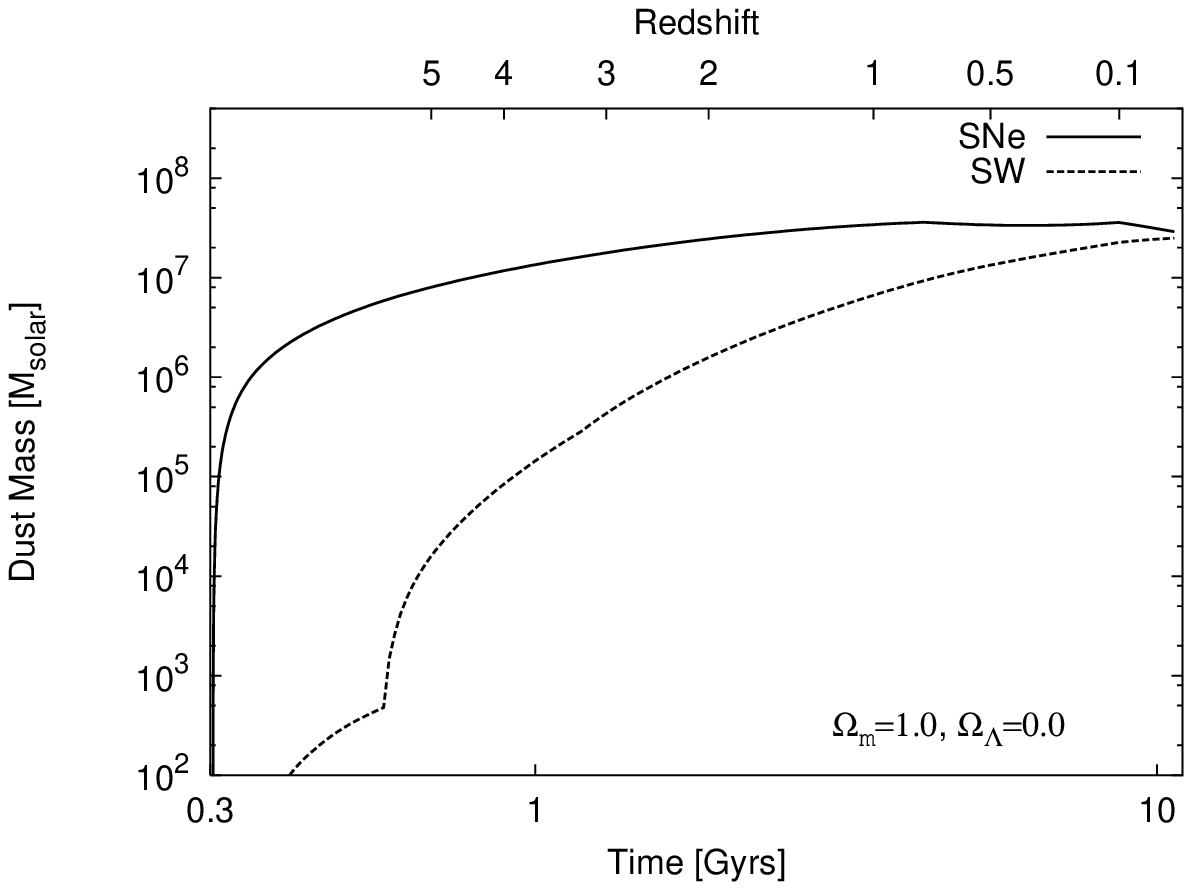}}
\epsfxsize=8cm
\subfigure[]{\epsffile{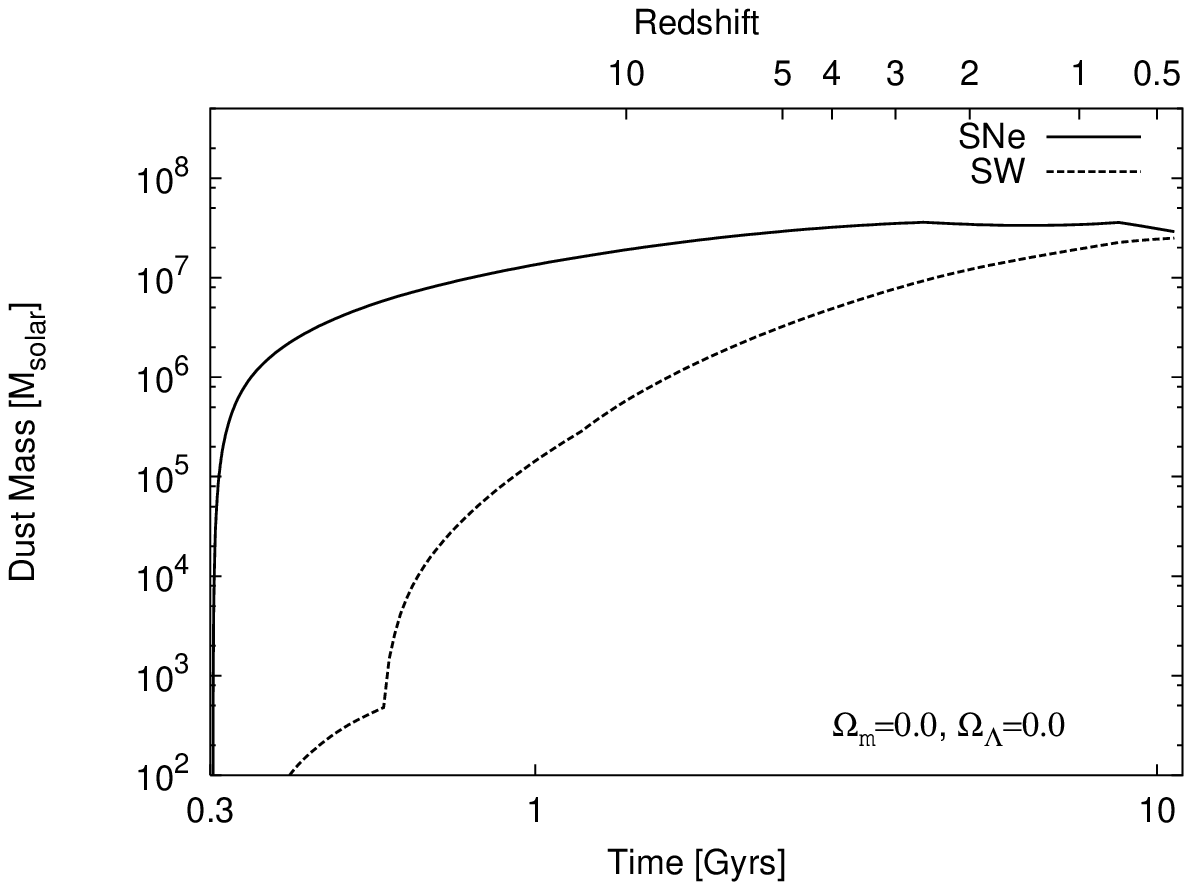}}
\epsfxsize=8cm
\subfigure[]{\epsffile{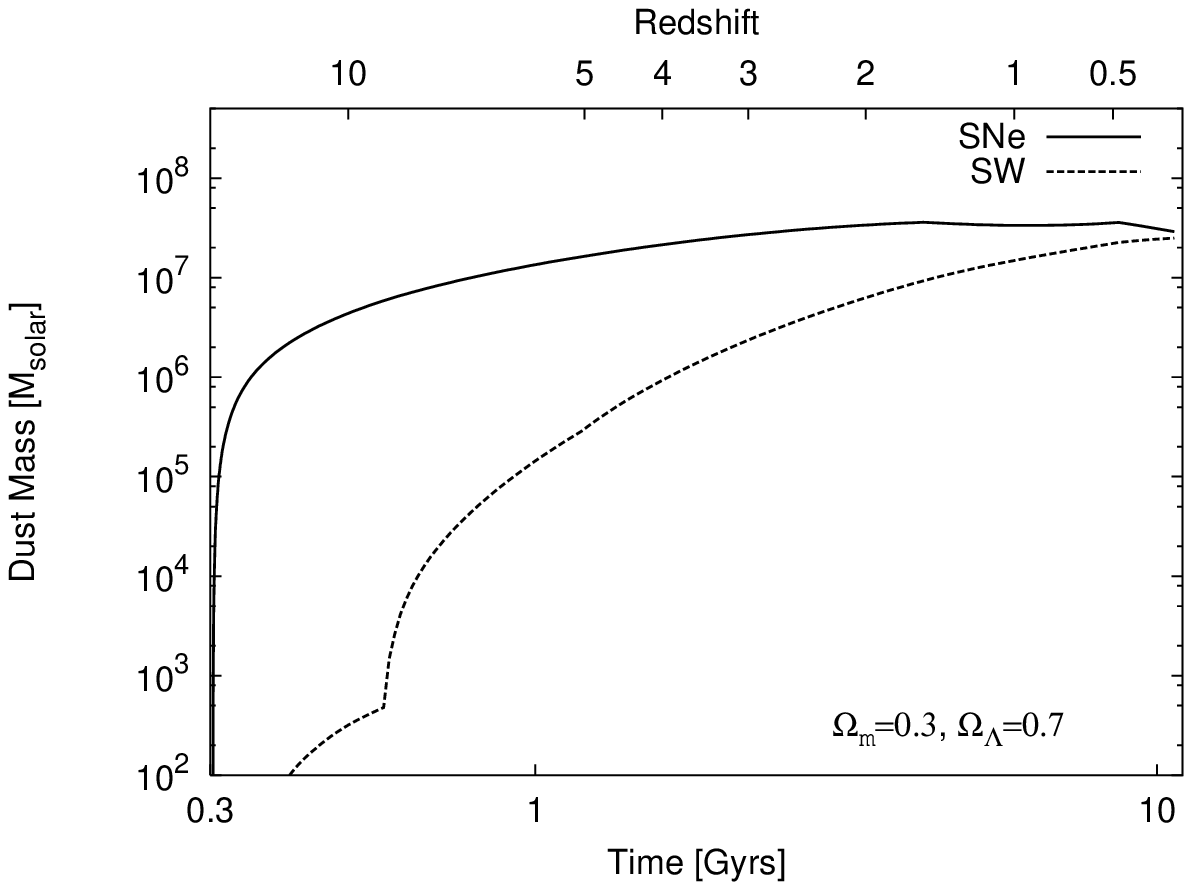}}
\epsfxsize=8cm
\subfigure[]{\epsffile{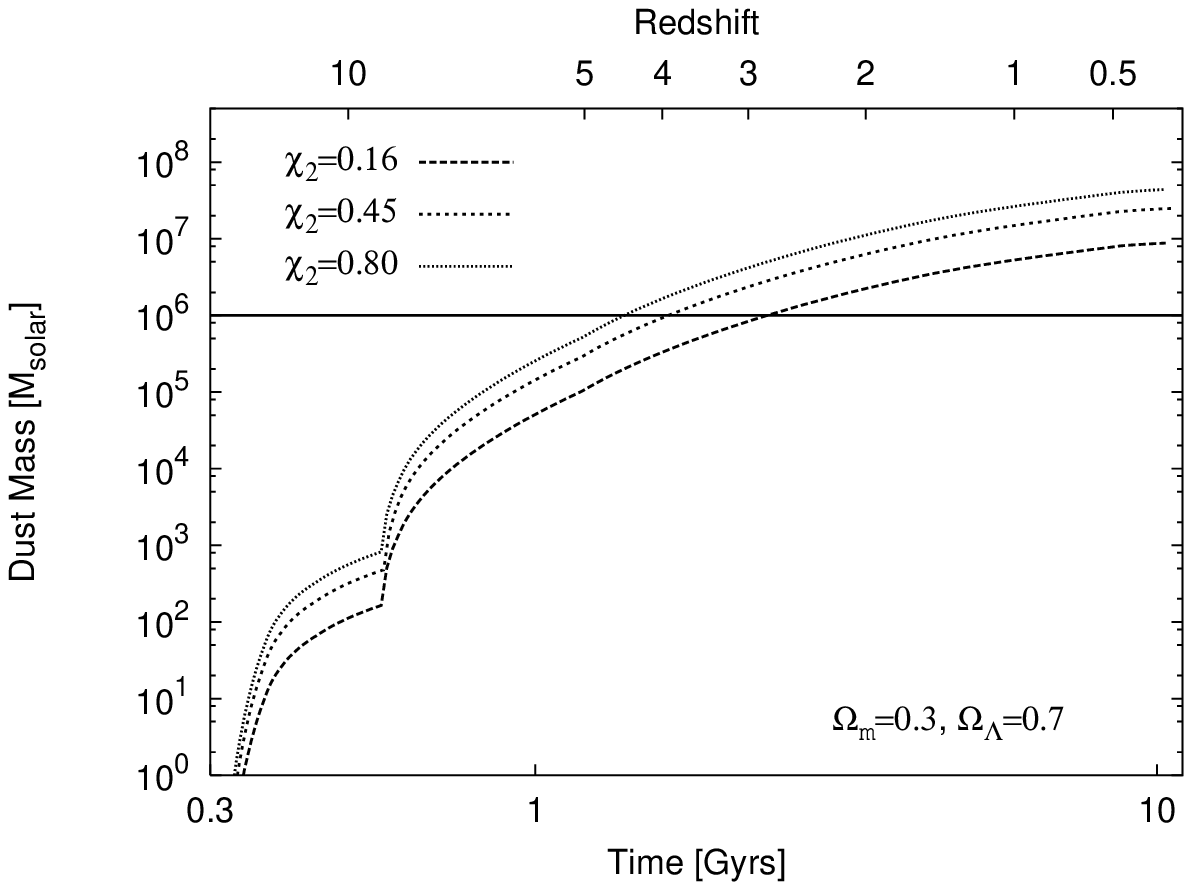}}
\caption{\small{The evolution of grain core mass yg ($M_{\odot}$) with
time and redshift for $g=2\times10^{10}M_{\odot}$, k=0.24.  The
cosmologies used are (a) $\Omega_m$=1, $\Omega_{\Lambda}$=0 (b)
$\Omega_m$=0, $\Omega_{\Lambda}$=0 and (c) $\Omega_m$=0.3,
$\Omega_{\Lambda}$=0.7 with star
formation 'turned on' at $t = 3\times10^{8}yrs$.  'SNe' and 'SW'
again represent the dust production in SNe and stellar winds with
values $\chi_1=0.23$, $\chi_2=0.45$.  (d) shows what would happen if we
take case (c) and alter the value of $\chi_2$ from 0.16, 0.45, 0.8.
The solid line represents 
the value of dust mass we have taken to be a significant amount and is 
a lower limit on the dust masses seen in high redshift SCUBA sources.}}
\label{red}
\end{figure*}
We can compare with another dust evolution model (Hirashita
et al, 2002) which represents a young starburst galaxy, SBS
{\small{0335-052}} with a star formation rate of $1M_{\odot}yr^{-1}$ 
(Figure~\ref{hirash} - note that this only shows the contribution from 
SNe dust). The model presented here 
predicts a similar increase in dust mass with time but is offset due
to a difference in the average $\chi_1$ term chosen
(taken conservatively in our case to be $\sim 0.2$ for $M < 40M_{\odot}$) and the yields
from massive stars (in this work taken from Maeder (1992)).   
An indicative dust mass derived from fitting a two component
temperature model to the IRAS and ISO observations of
SBS {\small{0335-052}} gives $M_{dust}\sim 2400M_{\odot}$ (Dale et al,
2001). The published errors in estimating the flux emitted and the temperatures
of the dust are $\sim 20-30$ per cent leading to large
uncertainties in the dust mass derived. As a guide, the model
presented here cannot produce this amount of dust until
$\sim10^7$yrs instead of $4$Myrs from the Hirashita model.  This is
still consistent with the observed age of the starburst modelled in
Hirashita et al (2002). 

Our predictions for an arbritrary galaxy with gas mass $g = 2\times
10^{10}M_{\odot}$ and star formation rate of $5M_{\odot}yr^{-1}$ ($k=0.24$) are
shown in this Section.  This represents galaxies similar to the Milky Way in which the gas 
mass fraction reaches 15 per cent after $\sim11$Gyrs.  The delay from long
lived stars returning dust to the ISM
is evident in Figure~\ref{delaychem} with significant dust production
only seen for $t>$ 2Gyrs.  Even if the condensation efficiency is
increased from 0.16 to 0.45 which is the maximum value of $\chi_2$ allowed
(i.e. 45 per cent of all returned metal mass in stellar winds is in
the form of dust), we still only have appreciable dust mass after 2
Gyrs.  Increasing the SFR rate by five times will result in a
higher dust mass at a given time, but the delay
from stars is still visible (Figure~\ref{k}).  SNe create more
dust from a higher SFR, but we still have the delay from SW's.
From Figures~\ref{delaychem} \&~\ref{k} we can clearly see that for
significant dust to
be available in metal poor, young galaxies it must come from SNe
unless star formation rates are very high.
This is further highlighted when mapping redshift onto time for
different cosmologies as seen in Figure~\ref{red}. 

To map redshift with time we need to know the cosmology i.e. the
values of $\Omega_m$ and $\Omega_{\Lambda}$.  Currently popular
values are 0.3 and 0.7 
respectively (i.e. a flat universe with nonzero cosmological
constant).  Other scenarios include $\Omega_m=1$
and $\Omega_{\Lambda}=0$ (again flat but with zero cosmological
constant) or $\Omega_m=0$ and $\Omega_{\Lambda}=0$ (i.e. an open
universe).  The age of the universe can be calculated from these
models using,
\begin{equation} 
t=H_o^{-1}\int_z^{\infty}{dz\over{(1+z)^{3/2}\sqrt{\Omega_m(1+z)^2+{\Omega_{\Lambda}}}}}
\label{timesteve}
\end{equation}
where $H_o$ is the Hubble constant (chosen here to be
$65kms^{-1}Mpc^{-1}$ - for a higher $H_o$ the corresponding time for a 
given redshift is lower) and for the simplest cases
Equation~\ref{timesteve} reduces to 
\[
\Omega_m, \Omega_{\Lambda} = 1,0 \Rightarrow  t={2/3H_o^{-1}\over{(1+z)^{3/2}}}
\]
and
\[
\Omega_m, \Omega_{\Lambda} = 0,0 \Rightarrow  t={H_o^{-1}\over{(1+z)}}
\]
We arbitrarily start star formation at $3\times 10^8$ years
corresponding to a redshift of 10 for 1, 0 cosmology and 14 for 0.3,
0.7 model.  It is believed that the first generation of star formation may
lie between z = 10 - 15 (Mackey et al, 2002) and it allows us to 'push' the time 
available for dust formation.  The evolution of grain core mass with time/redshift
starting from $t=3\times10^8$ years is plotted in Fig~\ref{red}.  These
figures clearly show that for the different
cosmologies given and a SFR efficiency of 0.24 (corresponding to a SFR of
$5M_{\odot}yr^{-1}$ for a $2\times10^{10}M_{\odot}$ galaxy), 
we require redshifts of less than 2, 8, or 4 respectively for dust
production in stellar winds to be
significant (i.e. when the dust mass fraction of the ISM (y) is
$\ge10^{-4}$ or alternatively when dust mass for the above system is $\ge
10^6M_{\odot}$).    Even if the
the condensation effciency in stellar winds is increased to 80 per
cent, we obtain similar results, Fig~\ref{red} (c).  The amount of dust mass produced
depends more on the coefficient of the star formation rate than
$\chi_2$ as shown in Figure~\ref{k}.
Indeed if the SFR efficiency is 2.5 ($50M_{\odot}yr^{-1}$ for this
system), dust production in winds reaches this level at redshifts less 
than 10, 30, 11 for the given cosmology (note: if $\chi_2$ {\em is}
actually 0.16 as proposed here then the ability to form dust is even harder -
significant dust production would require $z<3$ for $\Omega_m=0.3$,
and $\Omega_{\Lambda}=0.7$).
\\
\\
So if present cosmology is to be believed we may have a problem for the
creation of dust in these high redshift 'dusty' galaxies where z could 
be as large 10 (Eales et al, MNRAS submitted, 2002).  Dusty AGN are
certainly seen out to
redshifts $> 5$ (Robson et al, 2002).  If no dust is produced in SNe 
then to observe such galaxies we must set the condensation
efficency in winds at the maximum estimate ($\ge 0.45$) and have a high
star formation efficiency $k \ge 2.5$.
If this is not the case
then significant dust is not available until $z<5$.  If we
assume that dust is heated in regions of intense star formation by
massive stars then we may relate the observed FIR luminosity from high 
redshift sources to the current SFR of massive stars ($SFR
\propto L_{60}/L_{\odot}$).  
The flux densities reached in such observations suggest
that SFR's in the early Universe could be as high as
$1000M_{\odot}yr^{-1}$ (Gear et al, 2000).  With a galactic gas mass 
of order $10^{10}M_{\odot}$, a SFR as high as this could only sustain
itself for about $10^{7}$ years.  For the most massive galaxies
observed ($10^{12}M_{\odot}$), a SFR of $1000M_{\odot}yr^{-1}$
(e.g. Scott et al, 2002), represents a star
formation efficiency, k of $\sim 1$ and is equivalent to our nominal
galaxy with $g=2\times 10^{10}M_{\odot}$ and SFR =
$20M_{\odot}yr^{-1}$.  So although we have high SFR's inferred from
SCUBA sources, this is not enough to explain the dust seen in high
redshift galaxies, instead we require a high star formation {\it efficiency}.

\section{Lines of Sight}
\label{tau}
To determine the observational properties of these early
galaxies we need to know how soon they become optically thick due to the
dust that they are manufacturing.  Detailed modelling is beyond the
scope of the present work, and here we simply ask two questions.  The
first question is - what is the {\it average} optical depth 
seen when looking at the galaxy? i.e. would we be able, on average, to 
see right through the galaxy?  If the absorption is clumpy can we
still see deep into/through most of the galaxy?  This will have a
considerable influence on the galaxy's evolving optical and near-UV
appearance.  The second question is - what is the typical optical
depth to those parts of the galaxy where {\it most} of the radiated energy
is being generated? i.e. what is the optical depth to where young
massive stars are?  This will determine whether the bulk of energy is
actually re-radiated at far-infrared wavelengths.  Much fuller
methodS for considering escape and absorption of radiation in a clumpy
dusty environment are given by V\'{a}rosi \& Dwek (1999) and Gordon et 
al (2001), but for an
order-of-magnitude estimate here we just consider the optical depths
through a cloud of fixed mass, divided in various ways.  We start off
with a single (protogalactic) gas cloud of mass $M_o$, radius, $R_o$.
The optical depth through the cloud is given by
\begin{displaymath}
\tau_o ={\eta ZM_o\over{(4/3)^2\pi R_o^2r_d\rho_d}}
\end{displaymath}

For the closed box model, $Z = pln(1/f)$ with yield $p=0.02$ and fraction of
heavy elements incorporated into dust, $\eta=0.4$ ($r=0.1\mu m$,
$\rho_d = 2000kgm^{-3}$) we obtain, 
\begin{equation}
\tau_o = 15fln{1\over {f}}\left(M_{bo}\over{10^{11}M_{\odot}}\right)\left(R_o \over{10kpc}\right)^{-2}	
\label{taueq}
\end{equation}
For $f$ of order unity i.e. evolution of gas fraction from 1 to 0.8,
$fln1/f$ can be well approximated by $(1-f)$.  In Table~\ref{tautab} we
give some examples of the optical depth for a protogalactic cloud with 
mass $10^{11}M_{\odot}$ and radius 10kpc.
Equation~\ref{taueq} gives the optical depth for a galaxy regarded as
a single uniform cloud.  Suppose that we now divide this cloud up into N
smaller clouds each with mass $M_o/N$ and radius r.  The optical depth
of an {\it individual} cloud is now 
\begin{displaymath}
\tau = \tau_o\left(1 \over {N}\right)r^{-2}
\label{taucloud}
\end{displaymath}
The behaviour for different mass-radius relations of interstellar
clouds is interesting and
essentially well known, $M \propto r^{-\beta}$.  Uniform density
clouds would have $\beta = 3$ 
and the optical depth from Equation~\ref{taun} would {\it decrease} with
divison.  For clouds divided, but remaining in approximate
virial equilibrium, $\beta = 2$ (the Larson relation, e.g. see
references in Ashman \& Zepf, 2001) and the
optical depth of an individual cloud {\it does not change} with
division.  For clouds in free-fall (with a $density \propto r^{-2}$
structure), $\beta = 1$ and the optical depth of an individual cloud
{\it increases} with division.
\begin{table}
\centering
\caption{\label{tautab}\small{The optical depths of a protogalactic
cloud with $M_o=10^{11}M_{\odot}$ and $R_o = 10kpc$ for a given gas mass 
fraction, $f$ and metallicity, $Z$.}}
\begin{minipage}{8cm}
\centering
\begin{tabular}{cccccc} \hline
$f$ & 0.99 & 0.95 & 0.90 & 0.80 & 0.50\\ 
$Z/Z_{\odot}$ & 1/100 & 1/20 & 1/10 & 1/5 & 2/3\\
$\tau_o$ & 0.15 & 0.75 & 1.35 & 2.68 & 5.20\\ \hline
\end{tabular}
\end{minipage}
\end{table}
From the mass-radius relation, $r \propto M^{1/\beta}
\propto \left(M_o/N\right)^{1/\beta}$ hence
\begin{equation}
\tau = \tau_o\left(N^{2/\beta - 1}\right)
\label{taun}
\end{equation}

Now to answer the first question posed earlier - what is the mean optical
depth through the galaxy?  The answer is obvious, since the mass of
dust remains the same, but it is instructive to follow a simple
binomial distribution argument.  If the N clouds each have radius, $r$
and are randomly distributed through a galaxy of radius, $R_o$, then
(neglecting geometric factors) the chance of a line of sight through
the galaxy intercepting a particular cloud is 
\begin{equation}
p = {r^2 \over {R_o^2}}
\end{equation}
The probability of exactly n clouds being intercepted if there is a
total of N is the binomial distribution (with $q = 1-p$)
\[
^NC_n~p^nq^{N-n} = {N! \over {n!(N-n)!}}p^nq^{N-n}
\]
which has a $\mathit{mean} = Np$, and variance $Npq$.  The mean
optical depth, $\bar{\tau}$ is therefore $Np\tau$, 
\[
\bar{\tau} = Np\tau = N\left(1\over {N}\right)^{2/\beta}\tau_oN^{2/\beta - 1} 
= \tau_o
\]
as expected and {\it independent of the number of clouds}.  The
variance is $\tau_o \left(1-N^{-2/\beta} \right)$, tending to $\tau_o$ 
as N becomes large.

We can now sketch the early opactity evolution of a galaxy as it
fragments into star forming clouds.  The average optical depth through 
the galaxy will have a typical value in the visible of $\bar{\tau}
\leq 1$ until the metallicity has risen above $\sim 1/10$ solar
(Table~\ref{tautab}).  {\it The individual clouds however, will
rapidly become optically thick, even at abundances of 1/100 solar},
provided the mass-radius exponent $\beta$ is less than 2.  As an
example, at $Z \sim 1/100$ solar, if
$\beta \sim 1.5$ and the mass of a typical cloud is as high as $10^7$, then
N is of the order $10^4$ (Equation~\ref{taun}) and $\tau \sim 3$.  Star
formation is likely to occur in the centres of these clouds, so
extensive processing of radiation into the IR will take place if the
visual and UV optical depth out of the clouds is greater then one.
Most of the radiation will come from massive stars, and indeed McKee
\& Tan (2002) have suggested (in present day galaxies) that a mass
column density of order $1gcm^{-2}$ is needed for formation of the
massive stars.  The surface density of 
our fiducial cloud is given by
\begin{displaymath}
\sigma_g \approx 0.05f\left(M_o \over{10^{11}M_{\odot}}\right)\left(R_o \over{10kpc}\right)^{-2}~~~~gcm^{-2}
\end{displaymath}
so with $f \sim 1$, dividing into N clouds gives
\begin{displaymath}
\sigma_g \approx 0.05 \left(1\over{N}\right)N^{2/\beta} = 0.05N^{2/\beta - 1}~~~~~gcm^{-2}
\end{displaymath}

Evidently the suitable densities might never be reached if $\beta$ was 
equal to or around the value of 2, but with gravitational collapse in
the clouds its value will be lower and we can see that densities high
enough for massive clouds will
be approached, e.g. for $\beta = 1.5$ as soon as $N >
8000$ (corresponding to masses of order $10^8M_{\odot}$).  Such
clouds would become optically thick as soon as the metallicity reaches
above 1/100 solar, which answers our second question.

So we have a very elementary model of a gas cloud which fragments, forming
stars at the dense centers of the sub-clouds.  The optical depth from
the massive
stars outwards rapidly becomes large enough to reprocess the bulk of
radiation into the IR, even when the heavy element enrichment is low
(1/100 solar) - provided, of course, that about 40 per cent of the interstellar
metals are in the form of dust.  The average optical depth {\it
through} the galaxy can, however, remain low at $\tau_o \pm \tau_o$, where
$\tau_o \leq 1$ until the metallicity exceeds about 1/10 solar.

When will the optical depths decrease again?  A major process will be 
the dispersal of clouds due to the effects of massive stars
(e.g. Whitworth 1979, Matzner 2000) on timescales of $\sim$ a few
$10^7$ years.  Global winds may pull dust 
out of the galaxy and (perhaps relevant to elliptical galaxies)
$\tau_o$ will fall below 1 again, once $f$ is less than 0.02.

Overall, it is clear that the high IR and sub-mm flux from high
redshift star forming
galaxies is not suprising - there is ample dust available, provided
the incorporation of heavy elements into dust is a fairly rapid
process.  Either SNe are a major source of dust cores and/or star
formation rates are very high.

\section{Discussion and Conclusions.}
\label{sec:conc}
We conclude:

\begin{itemize}
\item  Published studies indicate the nucleation conditions for dust
formation are only reached in stars during their last (and shortest)
evolutionary stage (TPAGB for IMS and AGB/superwind phase for
$9-20M_{\odot}$).  It appears that it may be easier to form carbon
dust than silicates.

\item  In the study of dust in early galaxies a crucial unknown is the 
relative contribution of SNe and stellar winds.  SNe could inject dust
instantaneously into the ISM whereas SW's (unless from previously
unidentified massive star wind sources) must be delayed because of
chemical evolution and stellar lifetimes.  Galactic dust suggests
about equal contributions at the present day (E01).  By considering
current model atmospheres and stellar evolution tracks we conclude:
	\begin{itemize}
	\item  The normalised dust condensation efficiency in stellar winds
	can be set at a maximum of 45 per cent for solar metallicity,
	although we believe this may be much lower (around 16 per
	cent).  Dust production will be still lower at lower metallicities. 
	\item  In searching for dust formation sites it is likely that
	the TPAGB phase of IMS stars is the most promising site for dust
	formation in stellar winds.  It would be interesting to find if an
	equivalent stage actually exists for massive stars.
	\end{itemize}
\item  There is no difficulty in producing highly dusty galaxies at
redshifts above 5 if SNe are an important contributor to interstellar
dust.  If grain core destruction is mainly a result of SNe shocks the
process cannot be efficient if grains from the source are to survive.
We note that we have ignored dust destruction mechanisms in this
initial model, believing (E01) that such mechanisms may not be
efficient and require an extended timescale ($>10^9$yr) before they
can significantly alter the grain core content of the ISM.  Further
studies will be required to demonstrate this is indeed so.
\item If SNe are not a dominant dust source, significant dust masses
can only be generated at redshifts above 5 by galaxies with very high
star formation rates.
\item  High star formation rates required for significant dust
production at high redshifts are however consistent with the SFR
required to give the observed FIR luminosities, producing a consistent picture.
\item The average visual optical depth seen in a galaxy
with 40 per cent of the interstellar metals in the form of dust
will be less than 1 until a metallicity of $\sim 1/10$
solar is reached.  The optical depths seen in individual star forming
clouds can however reach values greater than 1 at very low metallicites
(1/100 solar), provided that the mass-radius exponent of
molecular clouds is less than 2.  The mass-radius
relation is instrumental in determining the optical depth observed in
high redshift galaxies if dust is present.  In a typical early
galaxy the visual optical depth through it would be patchy
($\tau_o \pm \tau_o$, with $\tau_o \sim$ a few), allowing glimpses of
emerging star forming regions, while most of the
radiation from the stars would emerge at IR wavelengths. 
\end{itemize}

In answer to the questions posed in Section~\ref{intro}, the source of dust at high
redshift may be attributed to stars only if there are high enough star
formation rates and condensation efficiencies.  It appears however,
that SNe dust must be required if dust is present at redshifts greater
than 10.  Whether this is an efficient process or not will be
addressed in future work.  Reduction of SCUBA data of young SNR's will
start shortly, with maps at 850 and 450$\mu m$
and available radio data providing a direct comparison between the
emission from dust and from any synchrotron emission from relativistic
electrons.  This work will
hopefully provide an estimate of an upper limit of any dust present
and a condensation efficiency of SNe.  Next steps include:
(1) using the modified elementary model to examine the composition and
size distribution of dust and its evolution (2) using the dust masses 
expected in early galaxies to examine the implications for molecular
formation onto grains in the ISM (3) theoretical work on dust
nucleation in SNe.  

We would encourage more extensive modelling of
dust formation processes particularly in stellar atmospheres with
inhomogeneities and in a wider variety of stars.

\section*{Acknowledgements}

HM would like to aknowledge a departmental grant from the  
University of Wales, Cardiff.  We would like to thank Loretta Dunne 
for invaluable comments on the draft manuscript and
Steve Eales and Tim Waskett for extremely
interesting and informative discussions.  We also thank the referee,
Eli Dwek for invaluable comments and his help in clarifying this work.

\end{document}